\documentclass[pre,twocolumn,superscriptaddress,showpacs]{revtex4}
\usepackage{graphicx}
\usepackage{amsmath}

\newcommand{\D}{\ensuremath{\mathrm{d}}}

\newcommand{\rme}{\ensuremath{\mathrm{e}}}

\DeclareMathOperator{\Order}{O}

\newcommand{\avg}[1]{\ensuremath{\langle #1 \rangle}}

\newcommand{\Dee}{\ensuremath{\hat{\mathcal{D}}}}
\newcommand{\Ell}{\ensuremath{\hat{\mathcal{L}}}}
\newcommand{\Lbias}[1]{\Ell^{\rm(bias)}_{#1}}
\newcommand{\Lint}[1]{\Ell^{\rm(int)}_{#1}}
\newcommand{\Lrep}[1]{\Ell^{\rm(rep)}_{#1}}

\begin{document}

\title{Utterance Selection Model of Language Change}
\date{\today}

\author{G.\ J.\ Baxter}
\affiliation{School of Physics and Astronomy, University of
Manchester, Manchester M13 9PL, U.K.}

\author{R.\ A.\ Blythe}
\affiliation{School of Physics and Astronomy, University of
Manchester, Manchester M13 9PL, U.K.}
\affiliation{School of Physics, University of Edinburgh, Mayfield
Road, Edinburgh EH9 3JZ, U.K.}

\author{W.\ Croft}
\affiliation{School of Languages, Linguistics and
Cultures, University of Manchester, Manchester M13 9PL, U.K.}

\author{A.\ J.\ McKane}
\affiliation{School of Physics and Astronomy, University of
Manchester, Manchester M13 9PL, U.K.}

\begin{abstract}
We present a mathematical formulation of a theory of language
change. The theory is evolutionary in nature and has close analogies
with theories of population genetics. The mathematical structure we
construct similarly has correspondences with the Fisher-Wright model
of population genetics, but there are significant differences. The
continuous time formulation of the model is expressed in terms of a
Fokker-Planck equation. This equation is exactly soluble in the case
of a single speaker and can be investigated analytically in the case
of multiple speakers who communicate equally with all other speakers
and give their utterances equal weight.  Whilst the stationary
properties of this system have much in common with the single-speaker
case, time-dependent properties are richer.  In the particular case
where linguistic forms can become extinct, we find that the presence
of many speakers causes a two-stage relaxation, the first being a common
marginal distribution that persists for a long time as a consequence
of ultimate extinction being due to rare fluctuations.
\end{abstract}


\maketitle

\section{Introduction}

Stochastic many-body processes have long been of interest to
physicists, largely from applications in condensed matter and chemical
physics, such as surface growth, the aggregation of structures,
reaction dynamics or pattern formation in systems far from equilibrium.   
Through these studies, statistical physicists have acquired a range of 
analytical and numerical techniques along with insights into the macroscopic 
phenomena that arise as a consequence of noise in the dynamics.  It is 
therefore not surprising that physicists have begun to use these methods 
to explore emergent phenomena in the wider class of \emph{complex} systems 
which---in addition to stochastic interactions---might invoke a 
\emph{selection} mechanism. In particular, this can lead to a system 
adapting to its environment.

The best-known process in which selection plays an important part is,
of course, biological evolution.  More generally, one can define an
evolutionary dynamics as being the interplay between three processes.
In addition to selection, one requires \emph{replication} (e.g., of
genes) to sustain a population and \emph{variation} (e.g., mutation)
so that there is something to select on.  A generalized evolutionary
theory has been formalized by biologist and philosopher of science
David Hull \cite{Hull88,Hull01} that includes as special cases both
biological and cultural evolution.  The latter of these describes, for
example, the propagation of ideas and theories through the scientific
community, with those theories that are ``fittest'' (perhaps by
predicting the widest range of experimental results) having a greater
chance of survival.  Within this generalized evolutionary framework, a
theory of language change has been developed
\cite{Croft00,Croft02,Croft05a} which we examine from the point of
view of statistical physics in this paper.

Since it is unlikely that the reader versed in statistical physics is
also an expert in linguistics, we spend some time in the next section
outlining this theory of language change.  Then, our formulation of a
very simple \emph{mathematical} model of language change that we
define in Sec.~\ref{moddef} should seem rather natural.  As this is
not the only evolutionary approach that has been taken to the problem
of language change, we provide---again, for the nonspecialist
reader---a brief overview of relevant modeling work one can find in
the literature.  The remainder of this paper is then devoted to a
mathematical analysis of our model.

A particular feature of this model is that all speakers continuously
vary their speech patterns according to utterances they hear from
other speakers.  Since in our model, the utterances produced represent
a finite-sized sample of an underlying distribution, the language
changes over time even in the absence of an explicit selection
mechanism.  This process is similar to \emph{genetic drift} that
occurs in biological populations when the individuals chosen to
produce offspring in the next generation are chosen entirely at
random. Our model also allows for language change by selection as well
as drift (see Sec.~\ref{moddef}). For this reason, we describe the
model as the ``utterance selection model'' \cite{Croft00}.

As it happens, the mathematics of our model of language change turn
out to be almost identical to those describing classical models in
population genetics.  This we discover from a Fokker-Planck equation
for the evolution of the language, the derivation of which is given in
Sec.~\ref{fpe}.  Consequently, we have surveyed the existing
literature on these models, and by doing so obtained a number of new
results which we outline in Sec.~\ref{singlespeaker} and whose
detailed derivation can be found elsewhere \cite{BBM05}.  Since in the
language context, these results pertain to the rather limiting case of
a single speaker---which is nevertheless nontrivial because speakers
monitor their own language use---we extend this in Sec.~\ref{multispeaker}
to a wider speech community.  In all cases we concentrate on
properties indicative of change, such as the probability that certain
forms of language fall into disuse, or the time it takes for them to
do so.  Establishing these basic facts is an important step towards
realizing our future aims of making a meaningful comparison with
observational data.  We outline such scope for future work and discuss
our results in the concluding section.

\section{Language change as an evolutionary process}
\label{framework}

In order to model language change we focus on \emph{linguistic
variables}, which are essentially ``different ways of saying the same
thing''.  Examples include the pronunciation of a vowel sound, or an
ordering of words according to their function in the sentence.  In
order to recognize change when it occurs, we will track the
frequencies with which distinct \emph{variants} of a particular
linguistic variable are reproduced in \emph{utterances} by a
language's \emph{speakers}.  Let us assume that amongst a given group
of speakers, one particular variant form is reproduced with a high
frequency.  This variant we shall refer to as the \emph{convention}
among that group of speakers.  Now, it may be that, over time, an
unconventional---possibly completely new---variant becomes more widely
used amongst this group of speakers.  Clearly one possibility here is
that by becoming the most frequently used variant, it is established
as the new convention at the expense of the existing one.  It is this
competition between variant forms, and particularly the propagation of
innovative forms across the speech community, that we are interested in.

We have so far two important ingredients in this picture of language
change: the speakers, and the utterances they produce.  The object
relating a speaker to her \footnote{We shall follow a convention where
speakers and hearers of a language are referred to using female and
male pronouns respectively.} utterances we call a \emph{grammar}.
More precisely, a speaker's grammar contains the entirety of her
knowledge of the language. We assume this to depend on the frequencies
she has heard particular variant forms used within her speech
community \cite{Bybee01, Pierrehumbert03}. In turn, grammars govern 
the variants that are uttered by
speakers, and how often.

Clearly, a ``real-world'' grammar must be an extremely complicated
object, encompassing a knowledge of many linguistic variables, their
variant forms and their suitability for a particular purpose.
However, it is noticed that even competent speakers (i.e., those who
are highly aware of the various conventions among different groups)
might use unconventional variants if they have become
\emph{entrenched} \cite{Croft00}.  For example, someone who has lived 
for a long time in one region may continue to use parts of the dialect 
of that region after moving to a completely new area.  This fact will 
impact on our modeling in two ways.  First, we shall assume that a 
given interaction (conversation) between two speakers has only a small 
effect on the established grammar.  Second, speakers will reinforce their 
own way of using language by keeping a record of their own utterances.

Another observed feature of language use is that there is considerable
variation, not just from speaker to speaker but also in the utterances
of a single speaker.  There are various proposals for the origin of
this variation.  On the one hand, there is evidence for certain
variants to be favored due to universal forces of language change.
For instance articulatory and acoustic properties of sounds, or
syntactic processing factors---which are presumed common to all
speakers---favor certain phonetic or syntactic changes over others 
\cite{Ohala83, Hawkins04}.
These universals can be recognized through a high frequency of such
changes occurring across many speech communities.

On the other hand, variation could reflect the wide range of possible
intentions a speaker could have in communicative enterprise.  For
example, a particular non-conventional choice of variant might arise
from the desire not to be misunderstood, or to impress, flatter or
amuse the listener \cite{Keller94}.  Nevertheless, in a recent
analysis of language use with a common goal \cite{Croft05b}, it was
observed that variation is present in nearly all utterances.  It seems
likely, therefore, that variation arises primarily as a consequence of
the fact that no two situations are exactly alike, nor do speakers 
construe a particular situation in exactly the same way. Hence there 
is a fundamental indeterminacy to the communicative process. As a 
result, speakers produce variant forms for the same meaning being 
communicated. These forms are words or constructions representing 
possibly novel combinations, and occasionally completely novel 
utterances.  Given the large number of possible sources of variation 
and innovation, we feel it appropriate to model these effects using 
a stochastic prescription.

In order to complete the evolutionary description, we require a
mechanism that selects an innovative variant for subsequent
propagation across the speech community.  In the theory of 
Ref.~\cite{Croft00} it is proposed that social forces play this role.  
This is based on the observation that speakers want to identify with 
certain subgroups of a society, and do so in part by preferentially 
producing the variants produced by members of the emulated subgroup 
\cite{Labov94,Milroy87}. That is, the preference of speakers to 
produce variants associated with certain social groups acts as a 
selection mechanism for those variants.

This particular evolutionary picture of language change (see
Sec.~\ref{compare} for contrasting approaches) places an emphasis on
utterances (perhaps more so than on the speakers).  Indeed, in
Ref.~\cite{Croft00} the utterance is taken as the linguistic analog of
DNA. As speakers reproduce utterances, linguistic structures get
passed on from generation to generation (which one might define as a
particular time interval).  For this reason, the term \emph{lingueme}
has been coined in \cite{Croft00} to refer to these structures, and to
emphasize the analogy with genetics.  One can then extend to analogy
to identify linguistic variables with a particular \emph{gene locus}
and variant forms with \emph{alleles}.

We stress, however, that the analogy between this evolutionary
formulation of language change and biological evolution is not exact.
The distinction is particularly clear when one views the two theories
in the more general framework of Hull \cite{Hull88,Hull01,Croft02}.
The two relevant concepts are \emph{interactors} and
\emph{replicators} whose roles are played in the biological system by
individual organisms and genes respectively.  In biology, a 
replicator (a gene) ``belongs to'' an interactor (an organism), 
thereby influencing its survival and reproductive ability of the interactor.  
This is then taken as the dominant force governing the make-up of the 
population of replicators in the next generation. The survivability of a 
replicator is not due to an inherent ``fitness'': it is the organism whose 
fitness leads to the differential survival or extinction of replicators. 
Also, the relationship between genotype and phenotype is indirect and complex.
Nevertheless, there is a sufficient correlation between genes and phenotypic 
traits of organisms such that the differential survival of the latter 
causes the differential survival of the former (this is Hull's definition of 
``selection''), but the correlation is not a simple one.

In the linguistic theory outlined here, the interactors (speakers) and 
replicators (linguemes) have quite different relationships to one another. 
The replicators are uttered by speakers, and there is no one-to-one 
relationship between a replicator (a lingueme) and the speaker who produces 
it. Nevertheless, Hull's generalized theory of selection can be applied 
to the lingueme as replicator and the speaker as interactor. 
Linguemes and lingueme variation is generated by speaker intercourse, 
just as new genotypes are generated by sexual intercourse. The 
generation process is replication, that is, speakers are replicating 
sounds, words and construction they have heard before. Finally, the 
differential survival of the speakers, that is, their social 
``success'', causes the differential survival of the linguemes they 
produce, and so the social mechanisms underlying the propagation of 
linguistic variants conforms to Hull's definition of selection.

In short, we do not suppose that the language uttered by an interactor
has any effect on its survival, believing the dominant effects on
language change to be social in origin.  That is, the survivability of
a replicator is not due to any inherent \emph{fitness}, but arises
instead from the social standing of individuals associated with the
use of the corresponding variant form. It is therefore necessary that
in formulating a mathematical model of language change, one should not
simply adapt an existing biological theory, but start from first
principles.  This is the program we now follow.

\section{Definition of the utterance selection model}
\label{moddef}

The utterance selection model comprises a set of rules that govern the
evolution of the simplest possible language viewed from the
perspective of the previous section.  This language has a single
lingueme with a restricted number $V \ge 2$ variant forms. At present
we simply assume the existence of multiple variants of a lingueme:
modeling the the communicative process and the means by which
indeterminacy in communication (see Sec.~\ref{framework}) leads to the
generation of variation is left for future work.

In the speech community we have $N$ individuals, each of whose
knowledge of the language---the grammar---is encoded in the set $X(t)$
of variables $x_{iv}(t)$.  In a manner shortly to be defined
precisely, the variable $x_{iv}(t)$ reflects speaker $i$'s ($1 \le i
\le N$) perception of the frequency with which lingueme variant $v$
($1 \le v \le V$) is used in the speech community at time $t$.  At all
times these variables are normalized so that the sum over all variants
for each speaker is unity, that is
\begin{equation}
\label{normalize}
\sum_{v=1}^{V} x_{iv}(t) = 1 \; \forall i, t \;.
\end{equation}
For convenience we will sometimes use a vector notation $\vec{x}_i = (
x_{i1}, \ldots, x_{iV} )$ to denote the entirety of speaker $i$'s
grammar.  The state of the system $X(t)$ at time $t$ is then the
aggregation of grammars $X(t) = ( \vec{x}_1(t), \ldots, \vec{x}_N(t))$.

After choosing some initial condition (e.g., a random initial
condition), we allow the system to evolve by repeatedly iterating the
following three steps in sequence, each iteration having duration
$\delta t$.

\begin{figure}
\includegraphics[width=0.7\linewidth]{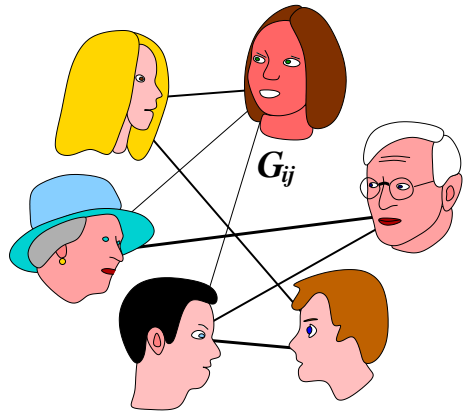}
\caption{\label{fig-society} Speakers in the society interact with
different frequencies (shown here schematically by different
thicknesses of lines connecting them).  The pair of speakers $i,j$ is 
chosen to interact with probability $G_{ij}$.}
\end{figure}

\emph{1. Social interaction.} A pair $i,j$ of speakers is chosen with
a (prescribed) probability $G_{ij}$.  There is no notion of an
ordering of a particular pair of speakers in this model, and so we
implicitly have $G_{ij} = G_{ji}$, normalized such that the sum over
\emph{distinct} pairs $\sum_{\langle i, j \rangle} G_{ij} = 1$.  See
Fig.~\ref{fig-society}.

\emph{2. Reproduction.} Both the speakers selected in step 1 produce a
set of $T$ \emph{tokens}, i.e., instances of lingueme variants.  Each
token is produced independently and at random, with the probability
that speaker $i$ utters variant $v$ equal to the \emph{production
probability} $x_{iv}^\prime(t)$ which will be determined in one of two
ways (see below).  The numbers of tokens $n_{i1}(t), \ldots,
n_{iV}(t)$ of each variant are then drawn from the multinomial
distribution
\begin{equation}
\label{multinomial}
P(\vec{n}_i , \vec{x}^{\,\prime}_i) = \binom{T}{n_{i1} \cdots n_{iV}}
(x^\prime_{i1})^{n_{i1}} \cdots (x^\prime_{iv})^{n_{iV}}
\end{equation}
where $\vec{x}^{\,\prime}_i = (x^{\prime}_{i1}, \ldots, x^{\prime}_{iV} )$,
$\vec{n}_i = (n_{i1}, \ldots, n_{iV} )$, $\sum_{v=1}^{V} n_{iv} = T$,  
and where we have dropped the explicit time dependence to lighten the 
notation.  Speaker $j$ produces a sequence of tokens according to the 
same prescription, with the obvious replacement $i \to j$.  The randomness 
in this step is intended to model the observed variation in language use 
that was described in the previous section.

\begin{figure}
\includegraphics[width=0.7\linewidth]{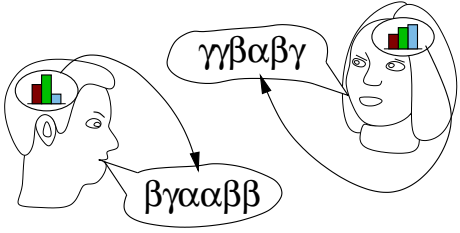}
\caption{\label{fig-unbiased} Both speakers $i$ and $j$ produce an
utterance, with particular lingueme variants appearing with a
frequency given by the value stored in the utterer's grammar when no
production biases are in operation. In this particular case three
variants are shown ($\alpha, \beta$ and $\gamma$) and the number of
tokens, $T$, is equal to 6.}
\end{figure}

The first, and simplest possible prescription for obtaining the
reproduction probabilities is simply to assign $x^\prime_{iv}(t) =
x_{iv}(t)$.  Since the grammar is a function of the speaker's
experience of language use---the next step explains precisely
how---this reproduction rule does not invoke any favoritism towards
any particular variants on behalf of the speaker.  We therefore refer
to this case as \emph{unbiased} reproduction, depicted in
Fig.~\ref{fig-unbiased}.

\begin{figure}
\includegraphics[width=0.7\linewidth]{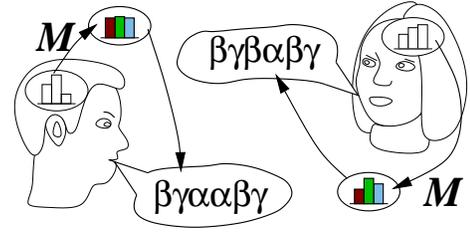}
\caption{\label{fig-biased} In the biased reproduction model, the
probability of uttering a particular variant is a linear combination
$M$ of the values stored in the grammar.}
\end{figure}

We shall also study a \emph{biased} reproduction model, illustrated in
Fig.~\ref{fig-biased}.  Here, the reproduction probabilities are a
linear transformation of the grammar frequencies, i.e.,
\begin{equation}
\label{lintrans}
x^\prime_{iv}(t) = \sum_{w} M_{vw} x_{iw}(t)
\end{equation}
in which the matrix $M$ must have column sums of unity so that the
production probabilities are properly normalized.  This matrix $M$ is
common to all speakers, which would be appropriate if one is
considering effects of universal forces (such as articulatory
considerations) on language.  Furthermore, in contrast to the unbiased
case, this reproduction model admits the possibility of innovation,
i.e., the production of variants that appear with zero frequency in a
speaker's grammar.

\begin{figure}
\includegraphics[width=0.7\linewidth]{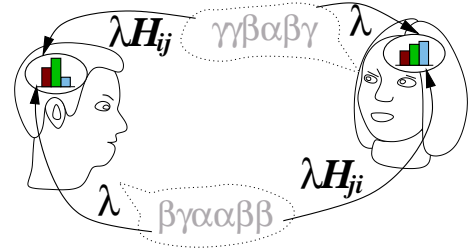}
\caption{\label{fig-retain} After the utterances have been produced,
both speakers modify their grammars by adding to them the frequencies
with which the variants were reproduced in the conversation.  Note
each speaker retains both her own utterances as well as those of her
interlocutor, albeit with different weights.}
\end{figure}

\emph{3. Retention.} The final step is to modify each speaker's grammar
to reflect the actual language used in the course of the interaction.
The simplest approach here is to add to the existing speaker's grammar 
additional contributions which reflect both the tokens produced by her
and by her interlocutor. The weight given to these tokens, relative to 
the existing grammar, is given by a parameter $\lambda$. Meanwhile, the 
weight, relative to her own utterances, that speaker $i$ gives to speaker 
$j$'s utterances is specified by $H_{ij}$.  This allows us to implement the 
social forces mentioned in the previous section. These considerations imply 
that
\begin{equation}
\label{update_nonorm}
\vec{x}_i (t + \delta t) \propto \left[ \vec{x}_{i}(t) + \lambda \left(
\frac{\vec{n}_{i}(t)}{T} + H_{ij} \frac{\vec{n}_{j}(t)}{T} \right) \right]
\end{equation}
for speaker $i$, and the same rule for speaker $j$ after exchanging all
$i$ and $j$ indices.  Fig.~\ref{fig-retain} illustrates this step. The 
parameter $\lambda$, which affects how much the grammar changes as a
result of the interaction is intended to be small, for reasons given in the 
previous section.  

We must also ensure that the normalization (\ref{normalize}) is maintained.  
Therefore,
\begin{equation}
\label{update}
\vec{x}_i (t + \delta t) = \frac{\vec{x}_{i}(t) + \frac{\lambda}{T} \left(
\vec{n}_{i}(t) + H_{ij} \vec{n}_{j}(t) \right)}{1 + \lambda(1+H_{ij})}\,.
\end{equation}

\begin{figure}
\includegraphics[width=0.7\linewidth]{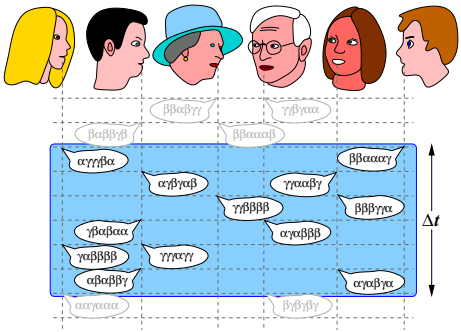}
\caption{\label{fig-generation} A generation of a population of
utterances in the utterance selection model could be defined as the 
set of tokens
produced by all speakers in the macroscopic time interval $\Delta t$.}
\end{figure}

Although we have couched this model in terms of the grammar variables
$x_{iv}(t)$, we should stress that these are not observable
quantities.  Really, we should think in terms of the population of
utterances produced in a particular generation, e.g., a time interval
$\Delta t \gg \delta t$ as indicated in Fig.~\ref{fig-generation}.
However, since the statistics of this population can be derived from the
grammar variables---indeed, in the absence of production biases they
are the same---we shall in the following focus on the latter.

\section{Comparison with other models of language change}
\label{compare}

Evolutionary modeling has a long history in the field of language
change and development.  Indeed, at a number of points in \emph{The
Origin of the Species}, Charles Darwin makes parallels between the
changes that occur in biological species and in languages.
Particularly, he used our everyday observation that languages tend to
change slowly and continuously over time to challenge the then
prevailing view that biological species were distinct species,
occupying immovable points in the space of all possible organisms.  As
evolutionary theories of biology have become more formalized, it is
not surprising that these there have been a number of attempts to
apply more formal evolutionary ideas to language change (see, e.g.,
\cite{Zuidema05}).  In this Section we describe a few of these studies
in order that the reader can see how our approach differs from others
one can find in the literature.

One area in which biological evolution plays a part is the development
of the capacity to use language (see, e.g., \cite{WM05} for a brief
overview).  Although this is in itself an interesting topic to study,
we do not suppose that this (presumably) genetic evolution is strongly
related to language change since the latter occurs on much shorter
timescales.  For example, the FOXP2 gene (which is believed to play a
role in both language production and comprehension) became fixed
around 120,000 years ago \cite{EPFetal02}, whereas the patterns in the
use of linguistic variables can change over periods as short as tens
of years.

Given an ability to use language, one can ask how the various
linguistic structures (such as particular aspects of grammar or
syntax) come into being \cite{CK03}.  Here evolutionary models that
place particular emphasis on language \emph{learning} are often
employed.  Some aspects of this type of work are reviewed in
\cite{Kirby02}---here we remark that in order to see the emergence of
grammatical rules, one must model a grammar at a much finer level than
we have done here.  Indeed, we have left aside the (nevertheless
interesting) question of \emph{how} an innovation is recognized as ``a
different way of saying the same thing'' by all speakers in the
community.  Instead, we assume that this agreement is always reached,
and concentrate on the fate of new variant forms.

Similar kinds of assumptions have been used in a learning-based
context by Niyogi and Berwick \cite{NB97} to study language change.
In learning-based models in general, the mechanism for language change
lies in speakers at an early stage of their life having a (usually
finite) set of possible grammars to choose from, and use the data
presented to them by other speakers to hypothesize the grammar being
used to generate utterances.  Since these data are finite, there is
the possibility for a child listening to language in use to infer a
grammar that differs from his parents', and becomes fixed once a
speaker reaches maturity.  Our model of continuous grammatical
change as a consequence of exposure to other speakers at all stages in
a speaker's life is quite different to learning-based approaches. In 
particular, it assumes an inductive model of language acquisition 
\cite{Tomasello03}, in which the child entertains hypotheses about 
sets of words and grammatical constructions rather than about entire 
discrete grammars. Thus, our model does not assume that a child has 
in her mind a large set of discrete grammars.

The specific model in \cite{NB97} assigns grammars (languages) to a
proportion of the population of speakers in a particular generation.
A particular learning algorithm then implies a mapping of the
proportions of speakers using a particular language from one
generation to the next.  Since one is dealing with nonlinear iterative
maps, one can find familiar phenomena such as bifurcations and phase
transitions \cite{Niyogi04} in the evolution of the language.  Note,
however, that the dynamics of the population of utterances and
speakers are essentially the same in this model, since the only thing
distinguishing speakers is grammar.  In the utterance selection model, we have
divorced the population dynamics of speakers and utterances, and allow
the former to be distinguished in terms of their social interactions
with other speakers (which is explicitly ignored in \cite{NB97}).
This has allowed us to take a \emph{fixed} population of
speakers without necessarily preventing the population of
utterances to change. In other words, language change may occur if 
the general structure of a society remains intact as individual 
speakers are replaced by their
offspring, or even during a period of time when there is no change in 
the makeup of the speaker population; both of these possibilities are 
widely observed.

An alternative approach to language change in the learning-based
tradition is not to have speakers attempt to infer the grammatical
rules underpinning their parents' language use, but to select a
grammar based on how well it permits them to communicate with other
members of the speech community.  This path has been followed most
notably by Nowak and coworkers in a series of papers (including
\cite{NKN01,KN03}) as well as by members of the statistical physics
community \cite{KKA05}.  This thinking allows one to borrow the notion
of \emph{fitness} from biological evolutionary theories---the more
people a particular grammar allows you to communicate with, the fitter
it is deemed to be.  In order for language use to change, speakers
using a more coherent grammar selectively produce more offspring than
others so that the language as a whole climbs a hill towards maximal
coherence.  The differences between this and our way of thinking
should be clear from Sec.~\ref{framework}.  In particular we assume no
connection between the language a speaker uses and her biological
reproductive fitness.  Finally on the subject of learning-based
models, we remark that not all of them assume language transmission
from parents to offspring.  For example, in \cite{MN04} the effects of
children also learning from their peers are investigated.

Perhaps closer in spirit to our own work are studies that have
languages competing for speakers.  The simplest model of this type is
due to Abrams and Strogatz \cite{AS03} which deems a language
``attractive'' if it is spoken by many speakers or has some
(prescribed) added value.  For example, one language might be of
greater use in a trading arrangement.  In \cite{AS03} good agreement
with available data for the number of speakers of minority languages
was found, revealing that the survival chances of such languages are
typically poor.  More recently, the model has been extended by Minett
and Wang \cite{MW04} to implement a structured society and the
possibility of bilingualism.  One might view the utterance selection
model as being relevant here if the variant forms of a lingueme
represent different languages.  However, there are then significant
differences in detail.  First, the way the utterance selection model
is set up would imply that all languages are mutually intelligible to
all speakers.  Second, in the models of \cite{AS03,MW04}, learning a
new language is a strategic decision whereas in the utterance
selection model it would occur simply through exposure to individuals
speaking that language.

To summarize, the distinctive feature of our modeling approach is that we
consider the dynamics of the population of utterances to be separate
from that of the speech community (if indeed the latter changes at
all).  Furthermore, we assume that language propagates purely through
exposure with social status being used as a selection process, rather
than through some property of the language itself such as coherence.
The purpose of this work is to establish an understanding of the
consequences of the assumptions we have made, particularly in those
cases where the utterance selection model can be solved exactly.

\section{Continuous-time limit and Fokker-Planck equation}
\label{fpe}

We begin our analysis of the utterance selection model by constructing a
Fokker-Planck equation via an appropriate continuous-time limit. There
are several ways one could proceed here.  For example, one could scale the
interaction probabilities $G_{ij}$ proportional to $\delta t$ (the
constant of proportionality then being an interaction rate).  Whilst
this would yield a perfectly acceptable continuous time process, the
Fokker-Planck equation that results is unwieldy and intractable. Therefore we
will not follow this path, but will discuss two other approaches below. The 
first will be applicable when the number of tokens is large. This will not 
generally be the case, but will serve to motivate the second approach, which 
is closer to the situation which we are modeling.

\subsection{The continuous time limit}
\label{continuous}
To clarify the derivation it is convenient to start with a single speaker
which, although linguistically trivial, is far from mathematically trivial. 
It also has an important correspondence to population dynamics, which is 
explored in more detail in Sec.~\ref{popgen}. In this case there is no matrix 
$H_{ij}$, and in fact we can drop the indices $i$ and $j$ completely. This 
means that the update rule (\ref{update}) takes the simpler form
\begin{equation}
\label{update_1}
\vec{x} (t + \delta t) = \frac{\vec{x} (t) + \frac{\lambda}{T} \vec{n} (t)}
{1 + \lambda}
\end{equation}
and so $\delta \vec{x} (t) \equiv \vec{x} (t + \delta t) - \vec{x} (t)$ is 
given by
\begin{equation}
\delta \vec{x} (t) = \frac{\lambda}{1+ \lambda} 
\left( \frac{\vec{n} (t)}{T} - \vec{x} (t) \right)\,.
\label{delta_x_1}
\end{equation}
The derivation of the Fokker-Planck equation involves the calculation of 
averages of powers of $\delta \vec{x} (t)$. Using Eq.~(\ref{multinomial}),
the average of $\vec{n}$ is $T \vec{x}^{\,\prime}$. If we begin by assuming
unbiased reproduction, then $\vec{x}^{\,\prime} = \vec{x}$ and so the average
of $\delta \vec{x} (t)$ is zero. In the language of stochastic dynamics, there
is no deterministic component --- the only contribution is from the diffusion
term. This is characterized by the second moment which is calculated in the 
Appendix to be
\begin{equation}
\langle \delta x_{v} (t) \delta x_{w} (t) \rangle = \frac{\lambda^2}
{(1+\lambda)^2} \frac{1}{T} \left( x_{v} \delta_{v w} - x_{v} x_{w}\right)\,,
\label{second_mom}
\end{equation}
where the angle brackets represent an average over all possible realizations.
To give a contribution to the Fokker-Planck equation, the second moment 
(\ref{second_mom}) has to be of order $\delta t$. One way to arrange this is 
as follows. We choose the unit of time such that $T$ utterances are made in 
unit time. Thus the time interval between utterances, $\delta t = 1/T$, is
small if $T$ is large. Furthermore, although the frequency of a particular 
variant in an utterance, $n_{v}/T$, varies in steps, the steps are very small. 
Therefore, when $T$ becomes very large, the time and variant frequency steps 
become very small and can be approximated as continuous variables. The second
jump moment, which is actually what appears in the Fokker-Planck equation,
is found by dividing the expression (\ref{second_mom}) by $\delta t = T^{-1}$,
and letting $\delta t \to 0$:
\begin{equation}
\alpha_{v w} (\vec{x}, t) = \frac{\lambda^2}{(1+\lambda)^2} 
\left( x_{v} \delta_{v w} - x_{v} x_{w}\right)\,.
\label{second_jump_mom}
\end{equation}
Since the higher moments of the multinomial distribution involve higher powers
of $T^{-1} = \delta t$, they give no contribution, and the only non-zero
jump moment is given by Eq.~(\ref{second_jump_mom}). As discussed in the 
Appendix, or in standard texts on the theory of stochastic processes
\cite{Risken89,Gardiner04}, this gives rise to the Fokker-Planck equation
\begin{equation}
\label{FPE_1_nobias}
\frac{\partial P(\vec{x}, t)}{\partial t} = \frac{\lambda^{2}}{2(1+\lambda)^2}
\sum_{v,w} \frac{\partial^{2} }{\partial x_v \partial x_w}
(x_v \delta_{v,w} - x_v x_w) P(\vec{x}, t)\,,
\end{equation}
where we have suppressed the dependence of the probability distribution 
function $P(\vec{x}, t)$ on the initial state of the system. 

The equation (\ref{FPE_1_nobias}) holds only for unbiased reproduction. It can 
be generalized to biased reproduction by noting that as $T \to \infty$, this
process becomes deterministic. Thus Eq.~(\ref{delta_x_1}) is replaced  by the
deterministic equation 
\begin{equation}
\delta \vec{x} = \frac{\lambda}{1+ \lambda} 
\left( \vec{x}^{\,\prime} - \vec{x} \right)\,.
\label{delta_x_1_deter}
\end{equation}
However, we may write Eq.~(\ref{lintrans}) using the condition 
$\sum_{w} M_{wv} = 1$ as
\begin{eqnarray}
x^\prime_{v} - x_{v} &=& \sum_{w} M_{vw} x_{w} - \sum_{w} M_{wv} x_{v} 
\nonumber \\
&=& \sum_{w \neq v} \left( M_{vw} x_{w} - M_{wv} x_{v} \right)\,.
\label{bias_deter}
\end{eqnarray}
The diagonal entries of $M$ are omitted in the last line because the condition 
$\sum_{w} M_{wv} = 1$ means that in each column one entry is not independent 
of the others. If we choose this entry to be the one with $w=v$, then all 
elements of $M$ in Eq.~(\ref{bias_deter}) are independent. Thus the diagonal 
entries of $M$ have no significance; they are simply given by 
$M_{vv} = 1 - \sum_{w \neq v} M_{wv}$. From Eqs.~(\ref{delta_x_1_deter}) and
(\ref{bias_deter}) we see that in order to obtain a finite limit as 
$\delta t \to 0$, we need to assume that the off-diagonal entries of $M$ are
of order $\delta t$. Specifically, we define $M_{v w} = m_{v w} \delta t$ for
$v \neq w$. Then in the limit $\delta t \to 0$,
\begin{equation}
\frac{d x_{v} (t)}{dt} = \frac{\lambda}{(1+\lambda)} \sum_{w \neq v} 
\left( m_{vw} x_{w} - m_{wv} x_{v} \right)\,.
\label{deter_eqn}
\end{equation}
Deterministic effects such as this give rise to ${\cal O} (\delta t)$ 
contributions in the derivation of the Fokker-Planck equation, unlike the
${\cal O} (\delta t)^{1/2}$ contributions arising from diffusion. Therefore,
the first jump moment in the case of biased reproduction is given by the 
right-hand side of Eq.~(\ref{deter_eqn}). The second jump moment is still 
given by Eq.~(\ref{second_jump_mom}), since any additional terms involving 
$M_{vw}$ are of order $\delta t$ and so give terms which vanish in the
$\delta t \to 0$ limit. This discussion may be straightforwardly extended to 
the case of many speakers. The only novel feature is the appearance of the 
matrix $H_{ij}$. In order to obtain a deterministic equation of the type 
(\ref{deter_eqn}), a new matrix has to be defined by 
$H_{ij} = h_{ij} \delta t$. 

Thus, in summary, what could be called the ``large $T$ approximation'' is 
obtained by choosing $\delta t = T^{-1}$, and defining new matrices $m$ and $h$
through $M_{v w} = m_{v w} \delta t$ for $v \neq w$ and 
$H_{ij} = h_{ij} \delta t$. It is the classic way of deriving the Fokker-Planck
equations as the ``diffusion approximation'' to a discrete process. However,
for our purposes it is not a very useful approximation. This is simply because 
we do not expect that in realistic situations the number of tokens will be 
large, so it would be useful to find another way of taking the continuous-time
limit. Fortunately, another parameter is present in the model which we have 
not yet utilized. This is $\lambda$, which characterizes the small effect 
that utterances have on the speaker's grammar. If we now return to the case
of a single speaker with unbiased reproduction, we see from 
Eq.~(\ref{second_mom}), that an alternative to taking $T^{-1} = \delta t$, is
to take $\lambda = (\delta t)^{1/2}$. Thus, in this second approach, we leave 
$T$ as a parameter in the model, and set the small parameter $\lambda$ equal 
to $(\delta t)^{1/2}$. The second jump moment (\ref{second_jump_mom}) in
this formulation is replaced by  
\begin{equation}
\alpha_{v w} (\vec{x}, t) = \frac{1}{T} 
\left( x_{v} \delta_{v w} - x_{v} x_{w}\right)\,.
\label{second_jump_mom_2}
\end{equation}
Bias may be introduced as before, and gives rise to 
Eqs.~(\ref{delta_x_1_deter}) and (\ref{bias_deter}). The difference in this
case is that $\lambda$ has been assumed to be ${\cal O} (\delta t)^{1/2}$,
and so the off-diagonal entries of $M$ (and the entries of $H$ in the case of 
more than one-speaker) have to be rescaled by $(\delta t)^{1/2}$, rather than 
$\delta t$. This means that in this second approach we must rescale the 
various parameters in the model according to
\begin{eqnarray}
\label{rescale1}
\lambda &=& (\delta t)^{1/2} \\
\label{rescale2}
M_{vw} &=& m_{vw} (\delta t)^{1/2} \quad\mbox{for}\quad v \ne w \\
H_{ij} &=& h_{ij} (\delta t)^{1/2}
\label{rescale3}
\end{eqnarray}
as $\delta t \to 0$.  We have found good agreement between the
predictions obtained using this continuous-time limit and the output
of Monte Carlo simulations when $\lambda$ was sufficiently small,
e.g., $\lambda \approx 10^{-3}$.

\subsection{The general form of the Fokker-Planck equation}
\label{gen_FPE}

In Sec.~\ref{continuous} we have outlined the considerations involved
in deriving a Fokker-Planck equation to describe the process. We concluded 
that, for our present purposes, the scalings given by 
Eqs.~(\ref{rescale1})-(\ref{rescale3}) were most appropriate. Much of 
the discussion was framed in terms of a single speaker, because the essential 
points are already present in this case, but here will study the full model. 
The resulting Fokker-Planck equation describes the time evolution of
the probability distribution function $P(X, t|X_{0}, 0)$ for the system to
be in state $X$ at time $t$ given it was originally in state $X_{0}$, although
we will frequently suppress the dependence on the initial conditions. The
variables $X$ comprise $N(V-1)$ \emph{independent} grammar variables, since
the grammar variable $x_{iV}$ is determined by the normalization 
$\sum_{v=1}^{V} x_{iv} = 1$.  

The derivation of the Fokker-Planck equation is given in the Appendix. It
contains three operators, each of which corresponds to a distinct dynamical
process. Specifically, one has for the evolution of the distribution
\begin{multline}
\label{driftfpe}
\frac{\partial P(X, t)}{\partial t} =
\sum_{i} G_i \left[ \Lbias{i} + \Lrep{i} \right] P(X, t) \\
{} + \sum_{\langle i j \rangle} G_{ij} \Lint{ij} P(X, t)
\end{multline}
in which $G_i = \sum_{j \ne i} G_{ij}$ is the probability that speaker
$i$ participates in any interaction.

The operator
\begin{equation}
\label{Lbias}
\Lbias{i} = \sum_{v=1}^{V-1} \frac{\partial}{\partial x_{iv}}
\sum_{\substack{w = 1 \\w \ne v}}^{V} (m_{w v} x_{i v} - m_{v w} x_{i
w})
\end{equation}
arises as a consequence of bias in the production probabilities.  Note
that the variable $x_{iV}$ appearing in this expression must be
replaced by $1 - \sum_{v=1}^{V-1} x_{iv}$ in order that the resulting
Fokker-Planck equation contains only the independent grammar variables.

As discussed above, the finite-size sampling of the (possibly biased)
production probabilities yields the stochastic contribution
\begin{equation}
\label{Lrep}
\Lrep{i} = \frac{1}{2T} \sum_{v=1}^{V-1} \sum_{w=1}^{V-1}
\frac{\partial^2}{\partial x_{iv} \partial x_{iw}} \left( x_{iv}
\delta_{v,w} - x_{iv} x_{iw} \right)
\end{equation}
to the Fokker-Planck equation.  In a physical interpretation, this
term describes for each speaker $i$ an independently diffusing
particle, albeit with a spatially-dependent diffusion constant, in the
$V{-}1$-dimensional space $0 \le x_{i1} + x_{i2} + \cdots + x_{i,V-1}
\le 1$.  On the boundaries of this space, one finds there is always a
zero eigenvalue of the diffusion matrix that corresponds to the
direction normal to the boundary.  This reflects the fact that, in the
absence of bias or interaction with other speakers, it is possible for
a variant to fall into disuse never to be uttered again.  These
\emph{extinction} events are of particular interest, and we
investigate them in more detail below.

The third and final contribution to (\ref{driftfpe}) comes from
speakers retaining a record of other's utterances.  This leads to
different speakers' grammars becoming coupled via the interaction term
\begin{equation}
\Lint{ij} = \sum_{v=1}^{V-1} \left( h_{i j} \frac{\partial}{\partial
x_{i v}} - h_{j i} \frac{\partial}{\partial x_{j v}} \right) \left(
x_{i v} - x_{j v} \right) \;.
\label{Lint}
\end{equation}

We end this section by rewriting the Fokker-Planck equation as a
continuity equation in the usual way: $\partial P/\partial t +
\sum_{i,v} \partial J_{i v}/\partial x_{i v} = 0$
\cite{Risken89,Gardiner04}, where
\begin{eqnarray}
J_{i v} (X, t) &=& - \sum_{\substack{w = 1 \\w \ne v}}^{V} 
G_{i} (m_{w v} x_{i v} - m_{v w} x_{iw}) P (X, t) \nonumber \\
&-& \frac{1}{2T} \sum_{w=1}^{V-1}
\frac{\partial}{\partial x_{iw}} G_{i} \left( x_{iv}
\delta_{v,w} - x_{iv} x_{iw} \right) P (X, t) \nonumber \\
&-& \sum_{j \neq i} G_{ij} h_{i j} \left( x_{i v} - x_{j v} \right) 
P (X, t) \;,
\label{current}
\end{eqnarray}
is the probability current. The boundary conditions on the Fokker-Planck 
equation with and without bias differ. In the former case, the boundaries are 
reflecting, that is, there is no probability current flowing through them. In
the latter case, they are so-called exit conditions: all the probability which 
diffuses to the boundary is extracted from the solution space. The result 
(\ref{current}) will be used in subsequent sections when finding the equations 
describing the time evolution of the moments of the probability distribution.

\section{Fisher-Wright population genetics}
\label{popgen}

The Fokker-Planck equation derived in the previous section is
well-known to population geneticists, being a continuous-time
description of simple models formulated in the 1930s by Fisher
\cite{Fisher30} and Wright \cite{Wright31}.  Despite criticism of
oversimplification (see, e.g., the short article by Crow \cite{Crow01}
for a brief history), these models have retained their status as
important paradigms of stochasticity in genetics to the present day.
Although biologists often discuss these models in the terms of
individuals that have two parents \cite{CK70,Burger00}, it is
sufficient for our purposes to describe the much simpler case of an
asexually reproducing population.

\begin{figure}
\includegraphics[width=\linewidth]{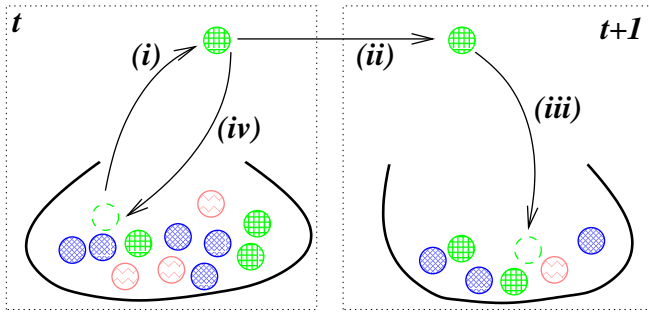}
\caption{\label{fig-beanbag} Fisher-Wright `beanbag' population
genetics.  The population in generation $t+1$ is constructed from
generation $t$ by (i) selecting a gene from the current
generation at random; (ii) copying this gene; (iii) placing the
copy in the next generation; (iv) returning the original to the
parent population.  These steps are repeated until generation
$t+1$ has the same sized population as generation $t$.}
\end{figure}

The central idea is that a given (integer) generation $t$ of the
population can be described in terms of a gene pool containing $K$
genes, of which a number $k_v$ have allele $A_v$ at a particular
locus, with $\sum^{V}_{v=1} k_{v} = V$ and $v=1, \ldots, V$. In the 
literature, an analogy with a bag containing $K$ beans is sometimes 
made, with different colored beans representing different alleles.  
The next generation is then formed by selecting \emph{with replacement} 
$K$ genes (beans) randomly from the current population.  This process 
is illustrated in Fig.~\ref{fig-beanbag}.  The replacement is crucial, 
since this allows for \emph{genetic drift}---i.e., changes in allele 
frequencies from one generation to the next from random sampling of 
parents---despite maintaining a fixed overall population size.

The probability of having $k^\prime_v$ copies of allele
$A_v$ in generation $t+1$, given that there were $k_v$ in the previous
generation, is easily shown to be multinomial, i.e.,
\begin{multline}
\label{FWmarkov}
P(k_1^\prime, k_2^\prime, \ldots, k_V^\prime; t+1 | k_1, k_2, \ldots, k_V; t)
= \\
\frac{K!}{k_1! k_2! \cdots k_V!} \left(\frac{k_1}{K}\right)^{k_1}
\left(\frac{k_2}{K}\right)^{k_2} \cdots
\left(\frac{k_V}{K}\right)^{k_V} \;.
\end{multline}
Using the properties of this distribution (see Appendix), it is
straightforward to learn that the mean change in the number of copies
of allele $A_v$ is the population from one generation to the next is
zero.  If we introduce $x_v(t)$ as the fraction $k_v/K$ of allele
$A_v$ in the gene pool at generation $t$, we find that the second
moment of this change is \cite{CK70}
\begin{multline}
\avg{[x_v(t+1) - x_v(t)] [x_w(t+1) - x_w(t)]} = \\
\frac{1}{2K} \left(
x_v(t) \delta_{v,w} - x_v(t) x_w(t) \right) \;.
\end{multline}
By following the procedure given in the Appendix, one obtains the
Fokker-Planck equation
\begin{equation}
\label{fwfpe}
\frac{\partial P(\vec{x}, t)}{\partial t} = \frac{1}{2K}
\sum_{v,w} \frac{\partial^{2} }{\partial x_v \partial x_w}
(x_v \delta_{v,w} - x_v x_w) P(\vec{x}, t)
\end{equation}
to leading order in $1/K$.  Since one is usually interested in large
populations, terms of higher order in $1/K$ that involve higher
derivatives are neglected.  Thus one obtains a continuous diffusion
equation for allele frequencies valid in the limit of a large (but
finite) population.

We see by comparing the right-hand side of (\ref{fwfpe}) with
(\ref{Lrep}) that the Fisher-Wright dynamics of allele frequencies in
a large biological population coincide with the stochastic component
of the evolution of a speaker's grammar.  Because of this mathematical
correspondence, it is useful occasionally to identify a speaker's
grammar with a biological population.  However, as noted at the end of
Sec.~\ref{moddef}, this should not be confused with the population of
utterances central in our approach to the problem of language change.

As we previously remarked, the fact that a speaker retains a record of
her own utterances means that the grammar of a single speaker will be
subject to drift, even in the absence of other speakers, or where zero
weight $H_{ij}$ given to other speaker's utterances.  In this case, a
single speaker's grammar exhibits essentially the same dynamics as a
biological population in the Fisher-Wright model.  We outline existing
results from the literature, as well as some extensions recently
obtained by us, in Sec.~\ref{singlespeaker} below.

The requirement that the population size $K$ is large for the validity
of the diffusion approximation (\ref{fwfpe}) of Fisher-Wright
population dynamics relates to the large-$T$ approximation of
Sec.~\ref{continuous}.  By contrast, the small-$\lambda$ approximation
relates to an \emph{ageing} population, i.e., one where a fraction
$\lambda/(1+\lambda)$ of the individuals are replaced in each
generation.  This is similar to a Moran model in population genetics
\cite{Moran58}, in which a single individual is replaced in each
generation.  Its continuous-time description is also given by
(\ref{fwfpe}) but with a modified effective population size $K$.

It is worth noting that when production biases are present, i.e., the
parameters $m_{vw}$ are nonzero, the resulting single-speaker
Fokker-Planck equation corresponds to a Fisher-Wright process in which
mutations occur \cite{CK70}.  In the beanbag picture, one would
realize this mutation by having a probability proportional to $m_{vw}$
of placing a bean of color $v$ in the next population, given that the
bean selected from the parent population was of color $w$.  It is
again possible to obtain exact results for this model, albeit for a
restricted set of mutation rates.  We discuss these below in
Sec.~\ref{singlespeaker}.

The remaining set of parameters in the utterance selection model,
$h_{ij}$, correspond to \emph{migration} rates from population $j$ to
$i$ in its biological interpretation.  It is apparently much more
difficult to treat populations coupled in this way under the
continuous-time diffusion approximation.  A prominent exception is
where one has two populations: a fixed mainland population and a
changing island population \cite{CK70}.  The assumption that the
mainland population is fixed is reasonable if it is much larger than
the island population.  Since a speaker's grammar does not have a
well-defined size, this way of thinking is unlikely to be of much
utility in the context of language change.  Therefore in
Sec.~\ref{multispeaker} we pursue the diffusion approximation where
all speakers (islands) are placed on the same footing.  This work
contrasts investigations based on ancestral lineages (``the
coalescent'') that one can find in the population genetics literature
(see, e.g., \cite{CCB03} for a recent review of applications to
geographically divided populations).  We shall also make use of these
results to gain an insight into the multi-speaker model.

Finally in this section we note that a feature ubiquitous in many
biological models, namely the selective advantage (or fitness) of
alleles, is not relevant in the context of language change.  For
reasons we have already discussed in Sec.~\ref{framework}, we do not
consider lingueme variants to possess any inherent fitness.

\section{Single-speaker model}
\label{singlespeaker}

We begin our analysis of the utterance selection model by considering
the case of a single speaker which is nontrivial because a speaker's
own utterances form part of the input to her own grammar.  We outline 
both relevant results that have been established in the population
genetics literature, along with an overview of our new findings which
we have presented in detail elsewhere \cite{BBM05}.  We begin with the
case where production biases (mutations) are absent.

\subsection{Unbiased production}

When the probability of uttering a particular variant form $v$ is
equal to the frequency $x_v$ stored in the speaker's grammar (we drop
the speaker subscript $i$ as there is only one of them), the
Fokker-Planck equation reads
\begin{equation}
\label{fpess}
\frac{\partial P(\vec{x}, t)}{\partial t} = \frac{1}{2T}
\sum_{v=1}^{V-1} \sum_{w=1}^{V-1} \frac{\partial^{2} }
{\partial x_v \partial x_w} (x_v \delta_{v,w} - x_v x_w) P
\end{equation}
where $V$ is the total number of possible variants.  We see that in
this case, $T$ enters only as a timescale and so we can put $T=1$ with
no loss of generality in the following.

One way to study the evolution of this system is through the
time-dependence of the moments of $x_v$.  Multiplying (\ref{fpess}) by
$x_v(t)^k$ and integrating by parts one finds \cite{BBM05}
\begin{equation}
\frac{\D \avg{ x_{v} (t)^{k}}}{\D t} = \frac{k(k-1)}{2} 
\left[ \avg{ x_{v} (t)^{(k-1})} - \avg{ x_{v} (t)^{k} } \right]\,.
\label{mom_eqn}
\end{equation}
We see immediately that the mean of $x_v$ is conserved by the
dynamics.  The higher moments have a time-dependence that can be
calculated iteratively for $k=2,3,\ldots$.  For example, for the
variance one finds that
\begin{equation}
\avg{x_v(t)^2} - \avg{x_v(t)}^2 = x_{v,0}(1-x_{v,0}) (1 -
\rme^{-t}) \;.
\end{equation}

Remarkably---and as we showed in \cite{BBM05}---the full
time-dependent solution of (\ref{fpess}) can be obtained under a
suitable change of variable.  The required transformation is
\begin{equation}
\label{xtou}
u_i = \frac{x_i}{1 - \sum_{j<i} x_j}
\end{equation}
which maps the space $0 \le x_1 + x_2 + \cdots + x_{V-1} \le 1$ onto
the $V{-}1$ dimensional unit hypercube, $0 \le u_i \le 1 \, \forall
i$. In the new coordinate system the Fokker-Planck equation is
\cite{BBM05}
\begin{eqnarray}
\label{fpuN1}
\frac{\partial P(\vec{u}, t)}{\partial t} &=& \frac{1}{2}
\sum_{v=1}^{V-1}
\frac{\partial^2}{\partial u_v^2} \frac{u_v(1-u_v)}{\prod_{w<v}
   (1-u_w)} P \\
&\equiv& \Dee_{V}(u_1, \ldots, u_{V-1}) P
\;.
\end{eqnarray}
The solution is then obtained by separation of variables.  First, we
separate the time and space variables so that given a fixed initial 
condition $\vec{u}_0$ one has
\begin{equation}
\label{modes}
P(\vec{u}, t) = \sum_{\lambda_V} C_{\lambda_V}(\vec{u}_0)
\Phi_{\lambda_V}(\vec{u}) \rme^{-\lambda_V t} \;.
\end{equation}
Here, $\lambda$ and $\Phi_{\lambda_V}(\vec{u})$ are the eigenvalues
and corresponding eigenfunctions of the operator $\Dee_V$ appearing in
(\ref{fpuN1}), and $C_{\lambda_V}(\vec{u}_0)$ a set of expansion
coefficients that are determined by the initial condition.

One can then separate each of the $u$ variables, since we have the recursion
\begin{equation}
\label{Drec}
\Dee_{V+1}(u_1, \ldots, u_V) = \Dee_2(u_1) + \frac{1}{1-u_1} 
\Dee_{V}(u_2, \ldots,
u_V) \;.
\end{equation}
To see this, let us assume we have found an eigenfunction
$\Phi_{\lambda_V}(u_1, \ldots, u_{V-1})$ of the $V$-variant operator
$\Dee_V(u_1, \ldots, u_{V-1})$ with accompanying eigenvalue
$\lambda_V$.  Now, we make an ansatz
\begin{equation}
\label{Phirec}
\Phi_{\lambda_{V+1}}(u_1, \ldots, u_V) = \psi_{\lambda_{V+1},
\lambda_{V}}(u_1) \Phi_{\lambda_{V}}(u_2, \ldots, u_V)
\end{equation}
for an eigenfunction of the $V{+}1$-variant operator $\Dee_{V+1}(u_1,
\ldots, u_V)$, where the corresponding eigenvalue $\lambda_{V+1}$
remains to be determined.  Inserting this ansatz into (\ref{Drec})
yields the ordinary differential equation
\begin{equation}
\frac{1}{2} \frac{\D^2}{\D u^2} u(1-u)
\psi_{\lambda_{V+1},\lambda_V}(u) = \left(\lambda_{V+1} - 
\frac{\lambda_V}{1-u} \right) \psi_{\lambda_{V+1},\lambda_V}(u)
\label{ode}
\end{equation}
that has to be solved for the function $\psi$.  Note that when $V=2$,
we have only one independent variable $u_1$ and the eigenfunction of
$\Dee_2(u_1)$ with eigenvalue $\lambda_2$ is the solution of
(\ref{ode}) with $\lambda_1=0$.  Beginning with this case in
(\ref{Phirec}) and iterating the requisite number of times, one finds
the solution for an eigenfunction of the $V$-variant Fokker-Planck
equation is
\begin{equation}
\label{eigey}
\Phi_{\lambda_{V}} = \psi_{\lambda_{V}, \lambda_{V-1}}(u_1)
\psi_{\lambda_{V-1}, \lambda_{V-2}}(u_2) \cdots \psi_{\lambda_2, \lambda_1}
(u_{V-1}) \;.
\end{equation}
That is, the partial differential equation (\ref{fpuN1}) is separable
in the variables $u_i$ as claimed, and each factor in the product is a
solution of the ordinary differential equation (\ref{ode}) that
contains two parameters.  After an appropriate substitution,
(\ref{ode}) can be brought into a standard hypergeometric form whose
solutions are Jacobi polynomials \cite{AS74}.  This analysis
\cite{BBM05} yields the eigenvalues of the Fokker-Planck equation.
When there are initially $V$ variants these are
\begin{equation}
\label{lambda}
\lambda_{V} = \frac{1}{2} L_{V-1} (L_{V-1} + 1) \,, \quad
L_v = \sum_{w=1}^{v} (\ell_w + 1)
\end{equation}
in which the variables $\ell_w$ are non-negative integers.  

\begin{figure}[htb]\begin{center}
\includegraphics[width=0.8\linewidth]{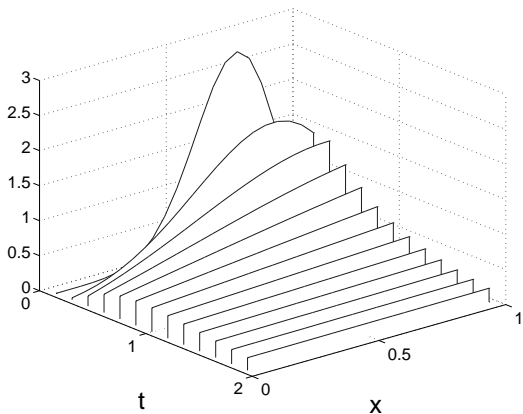}
\end{center}
\caption{\label{2var_sol_0}Time development of the exact solution of the
Fokker-Planck equation for a single speaker with two variants initially, 
when bias is absent and $x_0=0.7$.}
\end{figure}

Note that all the eigenvalues are positive: that is, the function
$P(\vec{u}, t)$ decays over time.  This is because of the fact that,
when no production biases are present, once a variant's frequency
vanishes, it can never be uttered again: i.e., variants become extinct
until eventually one of them becomes \emph{fixed}.  Hence, the
stationary probability distribution comprises delta functions at the
points where one of the frequencies $x_v=1$.  Since the mean of the
distribution is conserved (see above), the weight under each delta
function---which is the probability that the corresponding variant is
the only one in use as $t\to\infty$---is simply the variant's mean
frequency in the initial condition. Although we do not give the solution 
explicitly here, it is plotted for a two variant unbiased system in 
Figure~\ref{2var_sol_0}. The distribution in the interior of the domain 
decays with time, as the probability of one variant being eliminated (not 
plotted) grows.

It is remarkable that the solution of the Fokker-Planck equation for
$V$ variants is not much more complicated than the solution of the
corresponding equation for 2 variants. This turns out to be a feature
of other quantities associated with this problem. For example, the
probability $f_v(\vec{x}_0, t)$ that variant $v$ is the only one
remaining at a \emph{finite} time $t$, given an initial condition
$\vec{x}_0$, can be calculated rather easily because a reduction to an
effective two-variant problem can be found to work in this case as
well.  To understand this idea, it is helpful to return to the beanbag
picture of population genetics of the previous section.  We are
interested in knowing the probability that all beans in the bag have
the same color---say, for concreteness, chartreuse.  Let then $x$ be
the fraction of such beans in the bag in the current generation.  In
the next generation, each bean has a probability $x$ of being
chartreuse, and a probability $1-x$ of being some other color.
Clearly, the number of chartreuse beans in the next generation has the
distribution (\ref{FWmarkov}) with $V=2$, which is the reduction to the 
two-variant problem. The form of $f$ in this case was first found by 
Kimura \cite{Kimura55b} and is given by
\begin{multline}
\label{fv}
f_v(\vec{x}_0, t) = x_{v,0} -
\frac{1}{2} \sum_{\ell=1}^{\infty}
(-1)^{\ell} \big[ P_{\ell+1}(1-2x_{v,0}) - {}\\
P_{\ell-1}(1-2x_{v,0})  \big] \rme^{-\ell(\ell+1)t/2}
\end{multline}
in which $P_{\ell}(z)$ is a Legendre polynomial. Several other results can be
obtained by utilizing the above reduction to an equivalent two-variant problem
together with combinatorial arguments. For example, the probability that 
exactly $r$ variants coexist at time $t$ may be expressed entirely 
in terms of the function $f$ and various combinatorial factors
\cite{BBM05}.

Other quantities, such as the mean time to the $r$th extinction, or
the probability that a set of variants become extinct in a particular
order, can be most easily found from the backward Fokker-Planck
equation \cite{Risken89}, which involves the adjoint of the operator
$\Lrep{i}$.  In some cases, one can carry out a reduction to an 
equivalent two-variant problem wherein such quantities as the mean 
time to fixation of a variant $v$ averaged over those realizations 
of the dynamics in which it does become fixed \cite{KO69}
\begin{equation}
\label{tauv}
\tau_v = - 2 \frac{(1-x_{v,0}) \ln (1-x_{v,0})}{x_{v,0}}
\end{equation}
come into play.  Note, however, that this reduction is not always
possible.  For instance, in the two examples given at the start of
this paragraph, the former can be calculated from such a reduction,
whereas the latter cannot.  These subtleties are discussed in
\cite{BBM05}.

\subsection{Biased production}

We turn now to the case where the production probabilities and grammar
frequencies are not identical, but related by (\ref{lintrans}).  Here,
calculations analogous to those above are possible in those cases
where $m_{vw} = m_v$.  That is, in the interpretation where $m_{vw}$
are mutation rates, we can obtain solutions when mutation rates depend
only one the end product of the mutation.

To calculate moments of $x_v(t)$ it is most efficient to use the
Fokker-Planck equation in the form $\partial P/\partial t + \sum_{v}
\partial J_{v}/\partial x_{v} = 0$ and the explicit formula for the
current (\ref{current}) adapted to the single-speaker case to find the
equation satisfied by the moments:
\begin{eqnarray}
\frac{\D \avg{ x_{v} (t)^{k}}}{\D t} &=& \int \D\vec{x}\,x_{v}^{k}\, 
\frac{\partial P(\vec{x}, t)}{\partial t} \nonumber \\
= - \sum_{w} \int d\vec{x}\,x_{v}^{k}\,\frac{\partial J_{w}}{\partial x_w}
&=& k \int d\vec{x}\,x_{v}^{(k-1)}\,J_{v} (\vec{x}, t)\,, 
\label{continuity}
\end{eqnarray}
using the condition that the current vanishes on the boundary. Using
Eq.~(\ref{current}) the equation for the first moment, for instance,
is
\begin{eqnarray}
\frac{d \avg{ x_{v} (t) }}{dt} &=& - \sum_{w \neq v} \left( m_{w} \avg{ x_v }
- m_{v} \avg{ x_w } \right) \nonumber \\
&=& \left( - \sum_{w \neq v} m_{w} \right) \avg{ x_v} + m_{v} \left( 1 - 
\avg{ x_v } \right) \nonumber \\
&=& m_{v} - R \avg{ x_v }
\label{av_eqn}
\end{eqnarray}
in which $R=\sum_{v=1}^{V} m_v$.  This has the solution
\begin{equation}
\avg{x_v(t)} = \frac{m_v}{R} + \left( x_{v,0} - \frac{m_v}{R}
\right) \rme^{-R t} \;,
\label{av_soln}
\end{equation}
a result that does not depend on the number of tokens exchanged per
interaction since this affects only the stochastic part of the
evolution.  Higher moments have more complicated expressions which can
be found in \cite{BBM05}.

Once again, we can find the complete time-dependent solution of the
Fokker-Planck equation using the same change of variable and
separation of variables as before.  To achieve this, one makes the
replacement
\begin{equation}
\frac{1}{2} \frac{\partial}{\partial u_v} u_v(1-u_v) \to
\frac{1}{2T} \frac{\partial}{\partial u_v} u_v(1-u_v) + (R_v u_v -
m_v)
\end{equation}
in Eq.~(\ref{fpuN1}) and where we have introduced
\begin{equation}
R_v = \sum_{w=v}^{V} m_w \;.
\end{equation}
Note that it is necessary to reinstate the parameter $T$ since two
timescales are now in operation: one corresponding to the
probabilistic sampling effects, and the other to mutations.  In the
ensuing separation of variables, we find that each product $\psi$ in
the eigenfunction analogous to (\ref{eigey}) picks up a dependence on
the variant $v$ through the parameters $m_v$ and $R_v$.  The
eigenvalue spectrum also changes, becoming now
\begin{equation}
\lambda_V = \frac{1}{2T} L^\prime_{V-1}(2 T R + L^\prime_{M-1} - 1)
\,, \quad L^\prime_v = \sum_{w=1}^{v} \ell_w
\end{equation}
where $\ell_w$ are, as before, non-negative integers and
$R=\sum_{w=1}^{V} m_{w}$.  On this occasion, we have a zero eigenvalue
when $\ell_w = 0 \, \forall w$.  The corresponding eigenfunction is
then the the (unique) stationary state
$P^\ast(\vec{x})$ which is given by
\begin{equation}\label{SS_stat}
P^\ast(\vec{x}) = \Gamma(2R) \prod_{v=1}^{V} \frac{x_v^{2Tm_v -
1}}{\Gamma(2m_v)} \;.
\end{equation}
This result first appeared for the case $V=2$ in Ref.~\cite{Wright31}.

\begin{figure}[htb]\begin{center}
\includegraphics[width=0.7\linewidth]{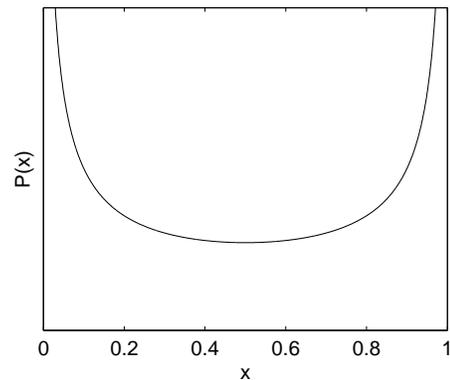}
\end{center}
\caption{The stationary distribution with one speaker and two variants for 
$m_1=m_2=0.2$.}\label{1sp_stationary20}
\end{figure}
\begin{figure}[htb]\begin{center}
\includegraphics[width=0.7\linewidth]{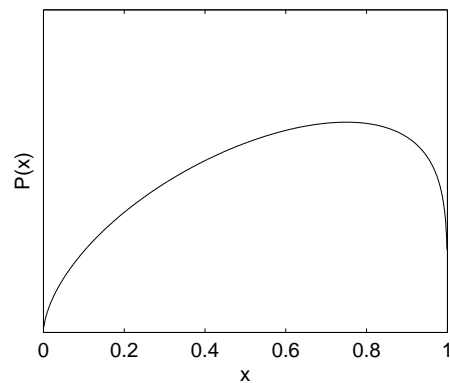}
\end{center}
\caption{The stationary distribution with one speaker and two variants for 
$m_1=0.8$ and $m_2=0.6$.}\label{1sp_stationary8060}
\end{figure}

When $V=2$, this is a beta distribution. It is peaked near the boundaries 
when $m_1$ and $m_2$ are both less than $1/2$, as illustrated in 
Figure~\ref{1sp_stationary20}. When the bias parameters are greater 
than $1/2$, the distribution is centrally peaked, and is asymmetric when 
$m_1\neq m_2$, as can be seen in Figure~\ref{1sp_stationary8060}.

It is perhaps interesting to note that the probability current is zero
everywhere in this steady state: i.e., that a detailed-balance
criterion is satisfied.  It seems likely that the more general
situation where $m_{vw}$ can depend both on the initial and final
variants will give rise to a steady state in which there is a
circulation of probability.  We believe a solution for this case has
not yet been found.

Finally in this survey of the single-speaker model we remark on the
existence of a hybrid model in which some of the production biases are
zero.  Then, those variants that have $x_v = 0$ will fall into disuse,
and the subsequent dynamics will be the same as for the case of biased
production among that subset of variants to which mutation is
possible.

\section{Multi-speaker model}
\label{multispeaker}

Having established the basic properties of the single speaker
model---moments, stationary distribution and fixation times---we now
seek their counterparts in the rather more realistic situation where
many different speakers are interacting.  The large number of
potential parameters specifying the interaction between speakers
($G_{ij}$ and $h_{ij}$) means the complexity of the multiple speaker
model is much greater than that for a single speaker.  However, some
analytic results can be obtained by considering the simplest set of
interactions between speakers, one where all the interaction
probabilities and weightings are equal. That is, we set
\begin{eqnarray}
\label{Ghflat}
G_{ij}\equiv G = \frac{1}{2N(N-1)} & \mbox{and} & h_{ij}\equiv h 
\;\;\forall \,i,j\;.
\end{eqnarray} 
This greatly simplifies the situation, as the interactions between speakers
are now identical, with different speakers being only distinguished by 
their initial conditions. From a linguistic point of view, it also seems 
natural to begin with all speakers interacting with the same probability,
as might happen in a small village \cite{Labov01,Trudgill04}. We are also
not considering social forces here, and so we assume that $H_{ij}$ is constant.
It can also be seen from the results for a single speaker that the majority 
of behaviors can be observed in systems with only two variants. Therefore we 
will not consider more than two variants for the remainder of this section. 

The Fokker-Planck equation (\ref{driftfpe}) now takes the relatively 
simple form
\begin{eqnarray}
\frac{\partial}{\partial t}P
& =(N-1)G\sum_i\left\{
\frac{\partial}{\partial x_i}(Rx_i-m_1)+
\frac{1}{2T}\frac{\partial^2}{\partial x_i^2}x_i(1-x_i)\right.\nonumber\\
& \left.+h\frac{\partial}{\partial x_i}(x_i-
\frac{1}{N-1}\sum_{j\neq i}x_j)\right\}P\nonumber\\
& =(N-1)G\sum_i\left\{
\frac{\partial}{\partial x_i}(Rx_i-m_1)+
\frac{1}{2T}\frac{\partial^2}{\partial x_i^2}x_i(1-x_i)\right.\nonumber\\
& \left.+\frac{N}{N-1}h\frac{\partial}{\partial x_i}(x_i-x)\right\}P
\label{flatFPE}
\end{eqnarray}
where we use $x$ without a subscript to denote the overall proportion of the 
first variant in the population $x \equiv \sum_ix_i/N$. The parameter $m_1$ 
is the bias parameter, $m_1\equiv m_{12}$, and
$R=m_1+m_2=m_{12}+m_{21}$.  Although we have not succeeded in solving
this equation exactly, we have been able to perform a number of
calculations and analyzes which we present below.

\subsection{Moments}
\label{ss_moments}

Differential equations for moments of $x_{i}$ can be found using the
same methods as before.  When production biases are present we find,
by multiplying (\ref{flatFPE}) by $x_i$, integrating and using the
fact that the probability current vanishes at the boundaries, that
(compare with Eq.~(\ref{av_eqn}))
\begin{equation}
\label{flat_mean_eq}
\frac{\D}{\D t}\langle x_i\rangle
=-(N-1)G\left[(R+h)\langle x_i\rangle 
-m_1-\frac{h}{N-1}\sum_{j\ne i} \avg{x_j}\right]\,.
\end{equation}
Note that the sum over $j$ in this expression can be written as
$N \langle x \rangle - \langle x_i \rangle$ where
\begin{equation}
\langle x \rangle = \frac{1}{N} \sum_i \langle x_i \rangle
\end{equation}
is the mean frequency over the entire community of speakers.

Using this substitution, and summing (\ref{flat_mean_eq}) over all
speakers, we find that
\begin{equation}
\frac{\D}{\D t} \langle x \rangle = - G (N-1) \left( R \langle x
\rangle - m_1 \right) \;.
\end{equation}
Subtracting this expression from (\ref{flat_mean_eq}) gives
\begin{equation}
\frac{\D}{\D t} \langle x_i - x \rangle = - G \left[ (N-1) R + N h
\right] \langle x_i - x \rangle \;.
\end{equation}
These equations are now decoupled and their solution follows readily
after implementing the initial condition and using the definitions
(\ref{Ghflat}).  We find that
\begin{eqnarray}
\langle x_i(t)\rangle &=& \frac{m_1}{R}+\left[(x_0-\frac{m_1}{R})\right.
\nonumber\\
&\,& +\left.(x_{i,0}-x_0)e^{-ht/2(N-1)}\right]e^{-Rt/2N}\label{mean_xi_t}\\
\langle x(t)\rangle &=&\frac{m_1}{R}+\left(x_0-\frac{m_1}{R}\right)e^{-Rt/2N}
\label{mean_x_t}\;,
\end{eqnarray}
where $x_0=x(0)=\frac{1}{N}\sum_i x_{i,0}$.

\begin{figure}[htb]\begin{center}
\includegraphics[width=\linewidth]{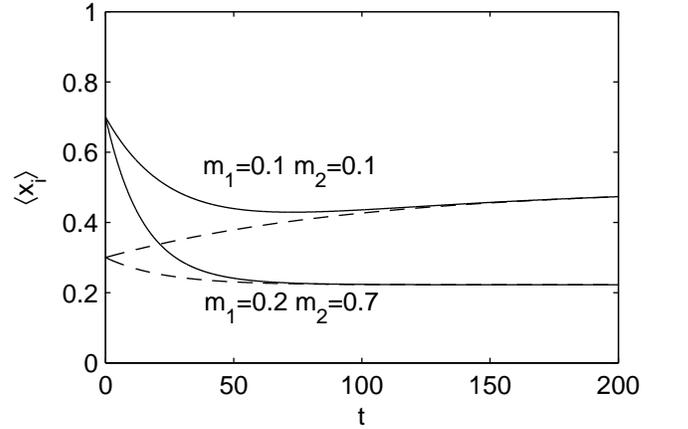}
\end{center}
\caption{The time development of the mean of a single speaker 
$\langle x_i\rangle$ for two different choices of mutation parameters. In 
each case $x_{i,0}=0.7$, $N=10$ and $h=0.5$. $T=1$. The overall population 
mean $\langle x\rangle$ is shown as a dashed line for comparison, with 
$x_0=0.3$.}\label{multi_mean_t}
\end{figure}

Each speaker's mean thus converges to the community's mean at a rate
controlled by $h$, and the latter relaxes to the fixed point of the
bias transformation $M$ at a rate determined by $R$.  In both cases,
the decay time grows linearly with the number of speakers $N$.  This
behavior is shown in Figure~\ref{multi_mean_t} in which the time
development of the mean of a particular speaker has been plotted for
two different bias parameter choices.

In the unbiased case we can repeat the same procedure to find the time
dependence of $\langle x_i\rangle$. The result is simply
(\ref{mean_xi_t}) and (\ref{mean_x_t}) with $R$ and $m_1$ set to zero,
though one must be careful with the boundaries when deriving the
equivalent of (\ref{flat_mean_eq}). In particular
\begin{equation}
\label{mean_res}
\langle x_i(t)\rangle=x_0+(x_{i,0}-x_0)e^{-ht/2(N-1)} \;,
\end{equation}
and we see explicitly that the expected overall fraction of each
variant in the population is conserved, just as in the single speaker
case:
\begin{equation}
\langle x(t)\rangle=x_0\;.
\end{equation}

Although we could write time dependent equations for higher moments, they are 
much more complicated.  Instead we now turn to the stationary
distribution.

\subsection{Stationary distribution}

In the absence of production biases, the stationary distribution is
one in which all speakers' grammars contain only one variant. This is
similar to the situation for a single speaker, only we should note
that (except in the special case of $h=0$, which is equivalent to the
single speaker case) equilibrium is only reached when \emph{all} the
speakers have the same variant.  Since $\langle x(t) \rangle$ is
conserved by the dynamics, we have once again that the weight under
the delta-function peaks equal to the initial mean frequency of the
corresponding variants within the entire community.  In the next
subsection, we shall investigate the relaxation to this absorbing
state of fixation.

When production biases are present, we expect an extended stationary
distribution with a mean given by (\ref{mean_x_t}) in the $t\to\infty$
limit. The second moments can be calculated by multiplying 
Eq.~(\ref{flatFPE}) by $x_i^2$ and $x_i x_j$, $i\neq j$, integrating, and 
using the fact that the probability current vanishes at the boundaries, just
as in the derivation of Eq.~(\ref{flat_mean_eq}), except that in this case 
there is no time derivative. Using the symmetry of the speakers we find that
\begin{align}
\left(R+h+\frac{1}{2T}\right)
\langle x_i^2\rangle^*-\left(m_1+\frac{1}{2T}\right)
\langle x_i\rangle^*-h\langle x_ix_j\rangle^*&=0\label{var_eqn}\\
\,[(N-1)R+h]  \langle x_ix_j\rangle^*-(N-1)m_1\langle x_i
\rangle^*-h\langle x_i^2\rangle^*&=0
\end{align}
where the asterisk denotes the steady state.  Solving gives
\begin{equation}
\langle x_i^2\rangle^*=\frac{m_1}{R}
\left\{\frac{(N-1)R\tilde{m}+h[(N-1)m_1+\tilde{m}]}{(N-1)R\tilde{R}+
h[(N-1)R+\tilde{R}]}\right\}\label{xi2_star}
\end{equation}
and, for $i \neq j$,
\begin{equation}
\langle x_ix_j\rangle^*=\frac{m_1}{R}
\left\{\frac{(N-1)m_1\tilde{R}+h[(N-1)m_1+\tilde{m}]}{(N-1)R\tilde{R}+h[(N-1)R+
\tilde{R}]}\right\}\label{xixj_star}
\end{equation}
where $\tilde{m}=m_1+1/2T$, $\tilde{R}=R+1/2T$. For the overall proportion 
of the first variant
\begin{align}
\langle x^2\rangle^*&=\frac{1}{N^2}\sum_{i,j}\langle x_ix_j\rangle^* 
\nonumber \\
&=\frac{m_1}{NR}\times \nonumber \\
&\!\!\!\!\!\!\!\!\!\!\left\{\frac{(N\!-\!1)R\tilde{m}+(N\!-\!1)^2m_1\tilde{R}+
Nh[(N\!-\!1)m_1+\tilde{m}]}{(N-1)R\tilde{R}+h[(N-1)R+\tilde{R}]}\right\}\,,
\end{align}
where the sum on the first line now includes the case $i=j$.

\begin{figure}
\includegraphics[width=0.8\linewidth]{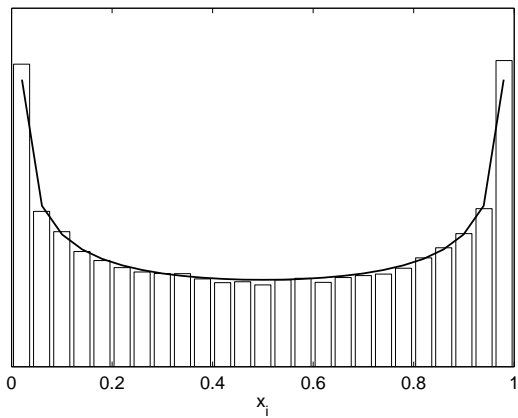}
\caption{\label{multi_stat} The single speaker marginal stationary 
distribution when $N=10$, $h=0.2\lambda$ and  $m_1=m_2=0.2$. Bars are the 
distribution obtained from simulation, while the curve is the approximate 
beta distribution. }
\end{figure}

When there are only two variants, the single speaker stationary
distribution (\ref{SS_stat}) is a beta distribution. The marginal
distribution for each speaker in the multiple speaker model is
modified by the presence of other speakers, but still the distribution
is peaked near the boundaries when the bias is small, and changes to a
centrally peaked distribution as the bias becomes stronger.  We
therefore propose that it is appropriate to approximate the stationary
marginal distribution as a beta distribution with mean and variance
just calculated.  That is,
\begin{equation}
P^*(x_i)\approx \frac{\Gamma(\alpha+\beta)}{\Gamma(\alpha)\Gamma(\beta)}
x_i^{\alpha-1}(1-x_i)^{\beta-1}\;,
\label{quasi}
\end{equation}
where
\begin{align}
\alpha &= 2Tm_1\!\left[\frac{(N-1)R+hN}{(N-1)R+h}\right]\;,\\
\beta &= 2T(R-m_1)\!\left[\frac{(N-1)R+hN}{(N-1)R+h}\right]\;.
\end{align}
Unlike in (\ref{SS_stat}) the parameters of the distribution now depend on  
$h$ and $N$ as well as $m_v$. The marginal distribution is well fitted by this 
beta distribution for a broad range of $h$ and $N$. An example is shown in 
Figure~\ref{multi_stat}, where the distribution calculated from simulations 
is compared to an approximating beta distribution. 

\begin{figure}[htb]\begin{center}
\includegraphics[width=0.95\linewidth]{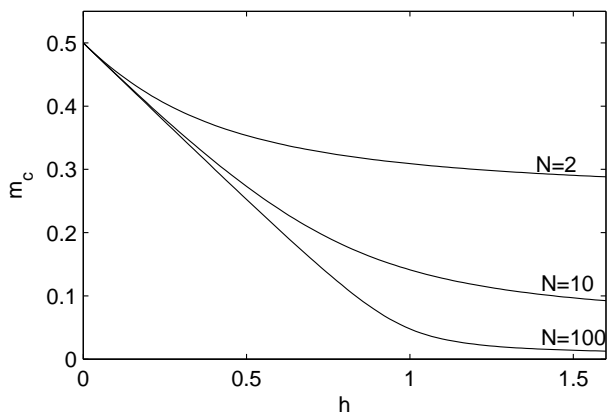}
\end{center}
\caption{The mutation value $m_c$ at which the stationary pdf of $x_i$ changes 
from a concave to a convex distribution, as a function of $h$ for 
$N=2$, $N=10$, $N=100$. Mutation is assumed symmetric: $m_1=m_2$.}
\label{multi_mtrans}
\end{figure}

When $N$ and $h$ are small, the transition from concave to convex shape occurs 
at approximately the same values of the mutation parameters as it does in the 
single speaker case, when $m_1=m_2=0.5$. As $N$ or $h$ become larger, the 
transition value becomes smaller. For sufficiently large $N$ or $h$, 
individual speakers will retain significant proportions of both variants, 
even for very small (but still non-zero) bias parameter values; the 
distribution will be centrally peaked unless $m_1$ and $m_2$ are extremely 
small. This can be seen in Figure~\ref{multi_mtrans}, which shows the value 
of $m=m_1=m_2$ at which the transition from concave to convex takes place for 
a range of $h$ and three different population sizes. This critical value of 
$m$, denoted by $m_c$, is the value of $m$ for which the parameters
$\alpha$ and $\beta$ in Eq.~(\ref{quasi}) pass through 1.

The stationary distribution of $x$ (the proportion of variant 1
throughout the population of speakers) on the other hand, does not
always have a simple shape. Consider first when the mutation strength
is fixed at some small value: $m_1=m_2\ll 0.5$. When $h$ is small some
speakers can be at opposite ends of the interval. For small $N$, this
leads to a multiply peaked distribution, with each peak representing a
certain fraction of the speakers being at one end. As $h$ gets larger,
the tendency to be at the same end increases, and the central peaks
dwindle, leaving the familiar double-peaked distribution. This only
holds so long as the mutation strength remains below the critical
value $m_c$, as shown in Figure~\ref{multi_mtrans}. For sufficiently
large $h$ or for larger $N$, the distribution becomes centrally
peaked.

\begin{figure}[htb]\begin{center}
\includegraphics[width=0.8\linewidth]{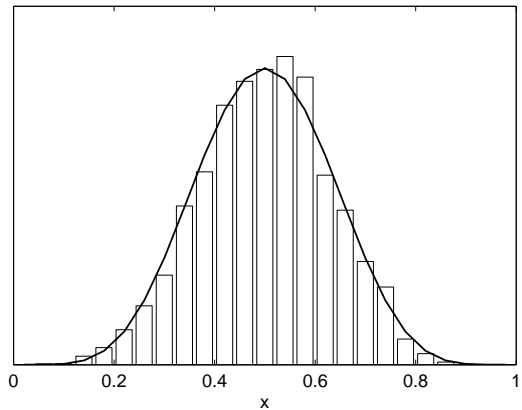}
\end{center}
\caption{\label{multi_stat2} The average speaker stationary 
distribution when $N=10$, $h=0.2\lambda$ and  $m_1=m_2=0.2$. Bars are the 
distribution obtained from simulation, while the curve is the approximate 
beta distribution.}
\end{figure}

When $m_1$ and $m_2$ are above the critical value, or if $N$ is
sufficiently large that the central-limit effect becomes significant,
the stationary distribution of $x$ is smooth and single peaked for all
values of $h$, becoming more bell shaped the higher the value of $N$
in accordance with the central limit theorem. Here we find that both
beta and Gaussian distributions calculated from the mean and second
moment fit well---see Figure~\ref{multi_stat2}. The value of $h$ only
has a small effect, altering the width of the distribution slightly.

\subsection{Fixation times}

\begin{figure}
\includegraphics[width=\linewidth]{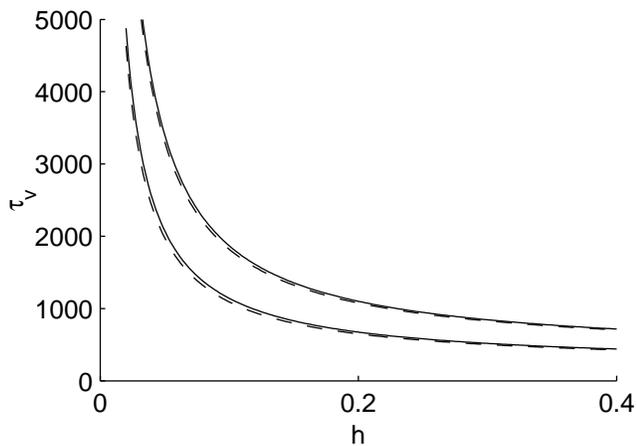}
\caption{\label{multi_fixtime} The mean time to fixation to each boundary as 
a function of $h$, for a system with 20 speakers and $x_0=0.3$. The solid 
curves are for an inhomogeneous initial condition, and the dashed curves are 
for a homogeneous initial condition. The lower curves are $\tau_2$ and the 
upper curves are $\tau_1$.}
\end{figure}

In the calculations of Sec.~\ref{ss_moments} we established that a single 
speaker's mean converges to the overall community's mean more slowly as 
the number of speakers is increased.  When production
biases are absent, we can also anticipate that the time to reach
fixation also increases with the number of speakers.  This fact can be
established analytically by re-casting the description of the system
in terms of the coalescent, a technique which can be found in
\cite{kin82,tak91}. We will not give the details of this calculation
here, but merely state the result, which is derived in
\cite{bly06}. The mean time to extinction of the second variant, which
corresponds to fixation of the first is
\begin{equation}
\tau_2[X(0)]=\frac{1-x_0}{x_0}\!\left[
\frac{N(N-1)}{2h}F[X(0)]-TN^2\ln(1-x_0)\right].
\end{equation}
Note that the second term is of the same form as (\ref{tauv}). The function 
$F$ depends on the initial distribution of speaker's grammars. For example, 
when all the speakers start with the same initial proportion 
($x_i(0)=x_0\;\forall i$),
\begin{equation}
F[X(0)]=\sum_{m=1}^{N-1}\frac{x_0^m}{m}-\frac{x_0}{N}
\frac{1-x_0^{N-1}}{1-x_0}\;,
\end{equation}
while when $M=Nx_0$ of the speakers start with $x_i=1$ and $N-M$ start 
with $x_i=0$ (so that the overall proportion is still the same),
\begin{equation}
F(X(0))=\sum_{m=1}^{M}\frac{\binom{M}{m}}{\binom{N}{m}}\frac{1}{m}\;.
\end{equation}
These are perhaps the extreme possibilities for the distribution, and in fact 
the values of $F$ differ little. For large $N$ they are virtually the 
same and both are well approximated by
\begin{equation}
F(X(0))\sim-\ln(1-x_0)
\end{equation}
which gives the much simpler expression for the mean time to extinction of the 
second variant
\begin{equation}\label{tau1_approx}
\tau_2\sim-\frac{1-x_0}{x_0}\ln(1-x_0)\left[
\frac{N(N-1)}{2h}+TN^2\right]
\end{equation}
that appeared in \cite{tak91}.  Figure~\ref{multi_fixtime} shows the
mean time to fixation at each boundary ($\tau_1$ and $\tau_2$) for a
system with only 20 speakers.  Already the times for inhomogeneous
(solid lines) and homogeneous (dashed lines) are very similar. Notice
also the dramatic increase in the fixation time as $h$ becomes
smaller. To calculate the mean to to fixation of \emph{any} variant,
we take a weighted average of the time for each variant:
\begin{align}
\tau &= x_0\tau_2+(1-x_0)\tau_1\nonumber\\
&\sim-\!\left[(1-x_0)\ln(1-x_0)\!+\!x_0\ln(x_0)\right]\!
\left[\frac{N(N-1)}{2h}+\!TN^2\right]\label{tau_total}.
\end{align}

\subsection{Quasi-stationary distribution en-route to fixation}

An interesting feature of the fixation time is that it increases
quadratically with the number of speakers $N$, whereas the moments
were seen to relax with time constants that grow linearly with $N$.
These results relate to the qualitative behavior observed in
simulation.  One notices the initial condition relaxes quickly to one
in which speakers have a distribution that persists for a long time
until a fluctuation causes extinction of a variant.  The nature of
this distribution depends on the size of $h$.  When it is very small,
the attraction of speakers to the boundaries is stronger than that to
the other speakers.  Therefore, some speakers dwell near the $x=0$
boundary, others near the $x=1$ boundary and only a few being in the
central part of the interval at any one time.  Here it is evident that
for fixation to occur, one needs all speakers near one of the
boundaries thus explaining why the fixation time is so much longer
than the initial relaxation.  For larger $h$, the attraction between
speakers overcomes the tendency to approach the boundaries, so the
speakers tend to dwell in the interior of the interval.

\begin{figure}[htb]
\includegraphics[width=0.8\linewidth]{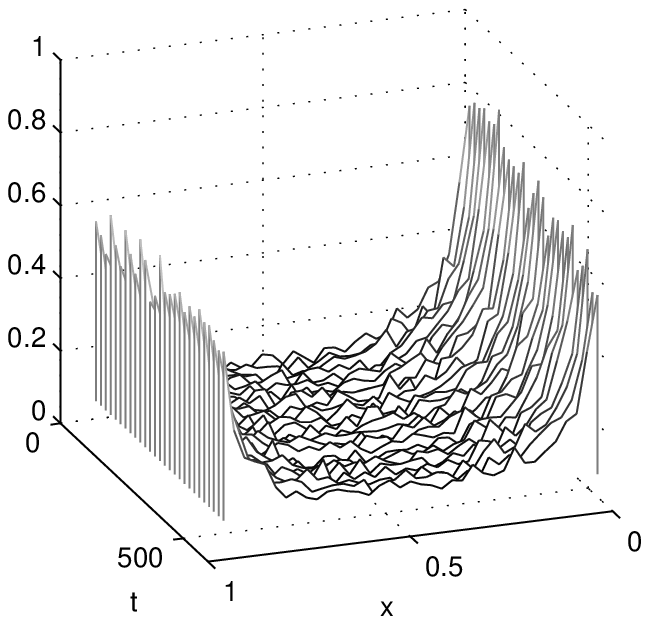}
\includegraphics[width=0.8\linewidth]{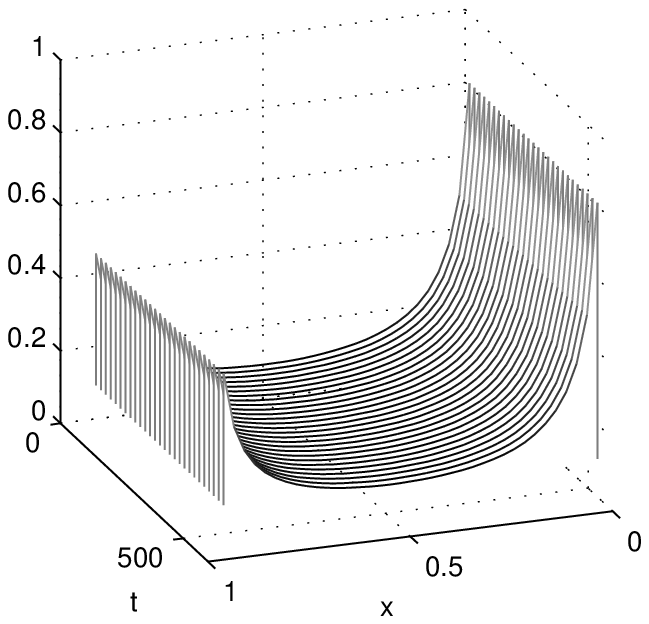}
\caption{\label{meta_dist}The distribution of speaker grammar values over a 
time series, for (top) an ensemble of realizations (none of which reach 
fixation during the period shown) and (bottom) the analytic beta distribution 
approximation, both for $N=20$ and $h=0.2\lambda$.}
\end{figure}

\begin{figure}
\includegraphics[width=0.9\linewidth]{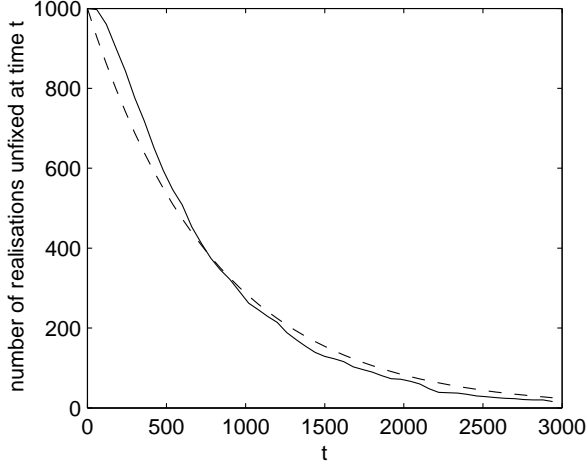}
\caption{\label{num_unfixed} The number of realizations remaining unfixed at 
time $t$, with initially 1000 realizations. Dashed curve is $1000e^{-t/\tau}$ 
where $\tau$ is given by Eq.~(\ref{tau_total})}
\end{figure}

We shall concentrate on the quasi-stationary distribution with $h$
small.  We obtain this using a mean-field argument, expected to be
valid for large $N$. As usual when applying mean-field theory we focus
on one constituent, in this case speaker $i$. We then replace the term
involving all the other speakers in the Fokker-Planck equation by an
average value.  Thus Eq.~(\ref{flatFPE}), in the unbiased case,
becomes
\begin{eqnarray}
\frac{\partial}{\partial t}P
& =(N-1)G\sum_{i} \left\{ \frac{1}{2T}\frac{\partial^2}{\partial x_i^2}
x_i(1-x_i)\right. \nonumber\\
& \left.+h\frac{\partial}{\partial x_i}(x_i- \avg{x})\right\}P\,.
\label{MF_FPE}
\end{eqnarray}
The solution to this equation is separable, so we write 
$P(X, t)=\prod_i p(x_{i}, t)$, and find the Fokker-Planck equation for
a single speaker to be 
\begin{eqnarray}
\frac{\partial}{\partial t} p(x_{i}, t)
& =(N-1)G\left\{\frac{\partial}{\partial x_i}(hx_i-h\avg{x})\right. \nonumber\\
& \left.+ \frac{1}{2T}\frac{\partial^2}{\partial x_i^2}x_i(1-x_i)\right\}
p(x_{i}, t)\,.
\label{singspeak_FPE}
\end{eqnarray}
After a rescaling of time $t \to (N-1)G t$, and dropping the index $i$, this
is exactly the Fokker-Planck equation for a single-speaker with bias and two
variants, with the identification $h \to R$ and $h\avg{x} \to m_1$. At 
large times we have from (\ref{mean_res}) that $\avg{x_{i}} = x_{0}=x_{i,0}$.
Therefore we expect that at large times the solution of the Fokker-Planck
equation to be identical to that of the single-speaker Fokker-Planck equation
with bias, as long as the identification $R \to h$ and $m_{1} \to hx_{0}$ is
made. In particular, we expect the marginal probability distribution for a
single speaker to have a stationary form which is a beta function of the
form (\ref{SS_stat}) with $V=2$ and $2Tm_{1} \to 2Thx_{0}$ and 
$2Tm_{2} \to 2Th(1-x_{0})$, that is,  
\begin{equation}
\label{ansatz_stat}
p_{\rm unfixed}(x_i)\sim \frac{\Gamma(\rho)}{\Gamma(\mu)\Gamma(\rho-\mu)}
x_i^{\mu-1}(1-x_i)^{(\rho-\mu)-1}\;,
\end{equation}
where
\begin{equation}
\rho=2Th \;\mbox{and}\; \mu=2Tx_0h\;.
\end{equation}
This distribution is shown in the lower half of Figure~\ref{meta_dist} 
for the case of $h$ small. In the upper half of this figure is the equivalent 
distribution calculated from numerical simulations, and it can be seen that 
the shape is maintained over time (the numerical result only includes 
realizations that do not fix in the time period specified), and that it is 
very similar to the beta approximation.

\begin{figure}[t]
\includegraphics[width=0.8\linewidth]{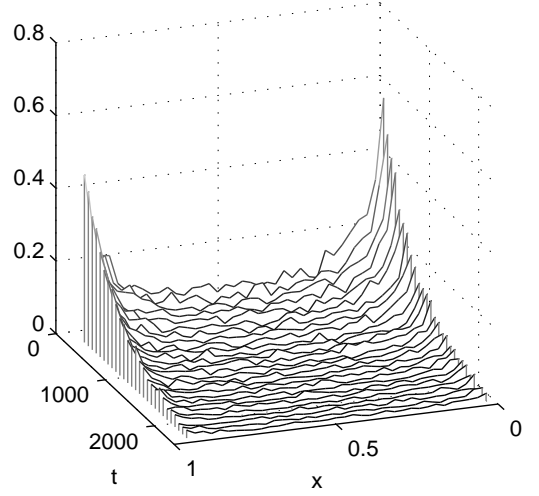}
\includegraphics[width=0.8\linewidth]{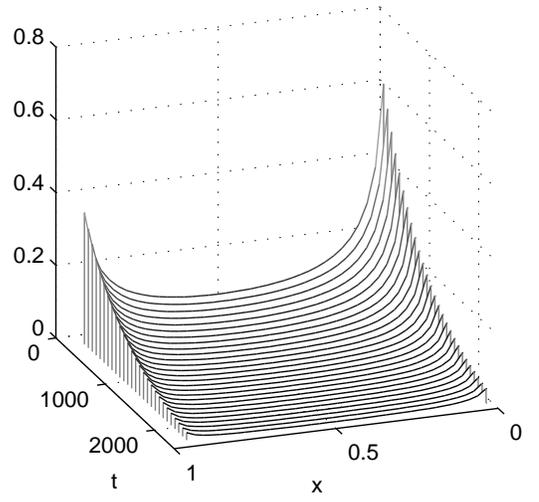}
\caption{\label{meta_timedev}The distribution of speaker grammar values over 
a time series, for (top) an ensemble of realizations (including fixing 
realizations) and (bottom) the analytic beta distribution approximation, both 
for $N=20$ and $h=0.2\lambda$.}
\end{figure}

If we assume that the rate at which any individual realization of the
process becomes fixed is constant, the number of unfixed realizations
exhibits an exponential decay with a time-constant $\tau$ given by
(\ref{tau_total}).  That this is the case is suggested by
Fig.~\ref{num_unfixed} in which the number of unfixed realizations as
a function of time obtained from Monte Carlo simulation is compared
with this prediction.  This then suggests for the full time-dependent
distribution the expression
\begin{equation}\label{ansatz_final}
p(x_i,t)\sim \frac{\Gamma(\rho)}{\Gamma(\mu)\Gamma(\rho-\mu)}
x_i^{\mu-1}(1-x_i)^{(\rho-\mu)-1}e^{-t/\tau}\;.
\end{equation}
In Figure~\ref{meta_timedev} we compare this approximation, shown in the lower
half, with numerical results in the upper half (where now the numerical 
results include realizations that fix during the time interval).  

\section{Discussion and Conclusion}
\label{conclusion}

In this paper we have cast a descriptive theory of language change,
developed by one of us \cite{Croft00,Croft02,Croft05a}, into a
mathematical form, specifically as a Markovian stochastic process. In
the resulting model there are a set of $N$ speakers who each have a
grammar which consists of $V$ possible variants of a particular
linguistic structure (a lingueme). In the initial phase of formulating
the process, two speakers out of the $N$ are picked out at every time
step and allowed to communicate with each other. The utterances they
produce modify the grammar of the other speaker --- as well as their
own --- by a small amount. Another two speakers are then picked at the
next time step and allowed to communicate. This process is repeated,
with two speakers $i$ and $j$ being chosen at each time step with a
probability $G_{ij}$. This matrix therefore prescribes the extent of
the social interaction between all speakers.

After many time steps the initial grammar of the speakers will have been 
modified in a way which depends on the choice of the model parameters. The
above formulation, that is, in terms of events which happen at regular 
time steps, is ideal for computer simulation. Of course, the model is 
stochastic, and so many independent runs have to be carried out, and the 
results obtained as averages over these runs. The randomness in the model
enters in two ways: in the choice of speakers $i$ and $j$ and in the choice 
of the variants spoken by a speaker in a particular utterance. We showed 
that it is possible to take the time interval between steps to zero, and 
derive a continuous time description of the process. When this procedure
is carried out, the model takes the form of a Fokker-Planck equation. 

The whole approach to language change we have been investigating was conceived 
as an evolutionary process, with linguemes being analogous to genes in 
population genetics. So it is perhaps not surprising that the mathematical
structures encountered when quantifying these theories are so similar. However,
as stressed in Sec.~\ref{popgen}, there are important differences. The most
direct correspondence with population genetics is when there is a single 
speaker and where the number of tokens is large. Furthermore, at each time step
the update rule (\ref{update_1}) applies in the linguistic model, whereas the 
equivalent rule in the population genetics case would be 
$\vec{x} (t + \delta t) = K^{-1} \vec{n} (t)$ corresponding to a completely 
new generation of $K$ individuals being created through random mating.
Thus the genetic counterpart is formally equivalent to letting 
$\lambda \to \infty$, and giving the previous grammar ($\vec{x} (t)$) zero
weight compared to the random element ($\vec{n} (t)$); for the actual 
linguistic problem, $\lambda$ is small, and it is $\vec{x} (t)$ that has by far
the greater weight. Taking $\lambda \to \infty$ and reinstating the factor of 
$T$ through a rescaling of the time, does indeed give the population genetics 
result (\ref{fwfpe}), with $K$ taking the role of $T$. Although the limit
$\lambda \to \infty$ is the precise correspondence, the scaling choice 
(\ref{rescale1})--(\ref{rescale3}) which we use also gives a mathematical,
if not a precise conceptual, equivalence between the genetic and linguistic
models.

Our analysis of the Fokker-Planck equation began by considering the case of 
only one speaker. This is far from trivial, and as we have seen is formally 
equivalent to standard models of population genetics. This has the advantage 
that many results from population genetics may be taken over essentially 
without change. Remarkably, the Fokker-Planck equation is in this case 
exactly soluble. This is due to the simple way in which the equation for $V$ 
variants is embedded in the $(V+1)$-variant equation. A similar simplification 
holds when calculating quantities such as the probability that a given number 
of variants coexist at time $t$ or the mean time to the $n$th extinction of 
a variant: they can be related by induction to the solution of the two-variant 
problem. 

While the exact solution of the mathematically non-trivial single
speaker case gives considerable insights into the effects caused by
the bias (or mutation) term (\ref{Lbias}) and the diffusion term
(\ref{Lrep}), to understand the evolution of variants across a speech
community it is clearly necessary to include the third term
(\ref{Lint}) in the Fokker-Planck equation. In Sec.~\ref{multispeaker}
we carried out an analysis of the model with this term included in the
simplest situation where all speakers were equally likely to talk to
all other speakers ($G_{ij}$ independent of $i$ and $j$) and where all
speakers gave the same weight to utterances from other speakers
($h_{ij}$ independent of $i$ and $j$). Just as for the single speaker
case, there are distinctions between the situations where there is
bias and where there is no bias. Whilst the presence of a bias
(through the term (\ref{Lbias})), makes the model more complicated,
its behavior is in fact simpler than if there were no bias: the
distribution of the probability of a variant in the population tends
to a stationary state which can be approximately characterized as a
beta distribution.  As we have seen, when no bias is present,
interactions between the speakers causes them all to converge
relatively quickly to a common marginal distribution which persists
for a long time until a fluctuation causes the same variant to be
fixed in all grammars.  Under a mean-field-type approximation, valid
in the limit of a large number of speakers, we established the form of
this quasi-stationary distribution.

In this paper, we have been primarily concerned with the mathematical
formulation of the theory and beginning a program of systematic
investigation of the model.  We believe that we have laid the
foundations for this study with the analysis we have presented, but
clearly there is much left to do.  In order to make connection with
observational data we will need to consider more realistic social
networks through which linguistic innovations may be
propagated---i.e., non-trivial $G_{ij}$, as in Fig.~1.  Bearing in
mind the proposed importance of social forces that described in
Sec.~\ref{framework}, it will also be necessary to include of speakers
or groups of speakers which may have more influence on language change
than others---i.e., non-trivial $H_{ij}$. Many of these cases will
only be amenable to analysis through computer simulations, but it
should be possible to obtain some analytical results with, for
example, a simplified network structure. However, it is clear that
even without any further developments, some of our results can be
generalized. For instance, by proceeding as in Sec.~\ref{ss_moments},
we can find that for general $G_{ij}$ and $h_{ij}$,
\begin{equation}
\frac{d \langle x_{i} \rangle}{dt} = \sum_{j \neq i} G_{ij} h_{ij} 
\left( \langle x_{i} \rangle - \langle x_{j} \rangle \right)\,,
\label{gen_1}
\end{equation}
and therefore that the rate of change of 
$\langle x \rangle = \sum_{i} \langle x_{i} \rangle$ is given by
\begin{equation}
\frac{d \langle x \rangle}{dt} = \sum_{i} \sum_{j \neq i} G_{ij} 
\left( h_{ij} - h_{ji} \right) \langle x_{i} \rangle\,.
\label{gen_2}
\end{equation}
Therefore $\langle x \rangle$ is conserved not only when $h$ is constant, as 
demonstrated in Sec.~\ref{ss_moments}, but also when $h_{ij}$ is symmetric.
In fact, the result can be further generalized. If we define the net ``rate of 
flow'' by 
\begin{equation}
\omega_{i} = \sum_{j \neq i} \left( G_{ij} h_{ij} - G_{ij} h_{ji} \right)\,,
\label{gen_3}
\end{equation}
then Eq.~(\ref{gen_2}) may be written as 
\begin{equation}
\frac{d \langle x \rangle}{dt} = \sum_{i} \omega_{i} \langle x_{i} \rangle\,.
\label{gen_4}
\end{equation}
So as long as $\omega_{i} = 0$ for all $i$, which may be thought of as
a kind of detailed balance condition, then the overall mean is
conserved. Now if the mean is conserved, then the probability of a
particular variant become fixed is simply its initial value. Therefore
no matter what the network or social structure, if $\sum_{j} G_{ij}
h_{ij} = \sum_{j} G_{ij} h_{ji}$ for all $i$, then this structure will
have no effect on the probability of fixation.

It is clear, however, that in general the further development of the
model will necessitate the choice of a particular network and social
structure. As an example of this we have recently begun to analyze the
model in the context of the formation of the New Zealand English
dialect, for which a reasonable amount of data is available
\cite{Gordon04,Trudgill04}.  In particular these give some information
about the frequencies with which different linguistic variables were
used by the first generations of native New Zealand English speakers
and their ultimate fate in the formation of today's conventional
dialect.  Predictions from our model relating to extinction
probabilities and timescales will play an important part in better
understanding this data.  More widely, we hope that the work presented
here will underpin many subsequent applications and form a basis for a
quantitative theory of language change.
 
\begin{acknowledgments}
RAB acknowledges both an EPSRC Fellowship (Grant GR/R44768) under
which this work was started, and a Scottish Executive/Royal Society of
Edinburgh Fellowship, under which it has been continued.  GJB thanks the 
NZ Tertiary Education Commission for a Top Achiever Doctoral Scholarship.
\end{acknowledgments}

\appendix
\section{Derivation of the Fokker-Planck equation}
\label{derivefpe}

In this Appendix we derive the Fokker-Planck equation (\ref{driftfpe}). The
method is standard, and involves the calculation of the so-called jump-moments
for the process under consideration \cite{Risken89,Gardiner04}. Since we
have already sketched some of the background in Sec.~\ref{continuous} for
the single speaker case, let us begin with this simpler version of the model.

Our starting point is the Kramers-Moyal expansion 
\begin{eqnarray}
\label{KM_1}
\frac{\partial P(\vec{x}, t)}{\partial t} &=& 
- \sum_{v=1}^{V-1} \frac{\partial}{\partial x_{v}} \left\{ \alpha_{v} (\vec{x})
P(\vec{x}, t) \right\} \nonumber \\
&+& \frac{1}{2} \sum_{v=1}^{V-1} \sum_{w=1}^{V-1} 
\frac{\partial^{2}}{\partial x_{v} \partial x_{w}} \left\{ \alpha_{v w} 
(\vec{x}) P(\vec{x}, t) \right\} \nonumber \\
&+& \ldots\,.
\end{eqnarray}
Here the dots represent higher order terms (which will turn out not to 
contribute) and the $\alpha$ functions are the jump moments 
\begin{eqnarray} 
\label{jump_mom_def1}
\alpha_{v} (\vec{x}) &=& \lim_{\delta t \to 0} \frac{\langle \delta x_{v} (t)
\rangle}{\delta t} \\
\alpha_{v w} (\vec{x}) &=& \lim_{\delta t \to 0} \frac{\langle \delta x_{v} (t)
\delta x_{w} (t) \rangle}{\delta t}\,,
\label{jump_mom_def2}
\end{eqnarray}
where $\delta x_{v} (t) \equiv x_{v} (t+\delta t) - x_{v} (t)$. The 
Kramers-Moyal expansion itself is derived from the assumption that the
stochastic process is Markov together with a continuous time approximation
\cite{Risken89,Gardiner04}.

In the single speaker case we have already established a form for 
$\delta x_{v} (t) $ (see Eq.~(\ref{delta_x_1})) and since the mean of the 
multinomial distribution (\ref{multinomial}) is simply
\begin{equation}
\label{navg_1}
\avg{n_{v}} = T x_{v}^\prime \;,
\end{equation}
a manipulation as in Eq.~(\ref{bias_deter}) and a rescaling as in 
Eqs.~(\ref{rescale1}) and (\ref{rescale2}) leads to
\begin{equation}
\langle \delta x_{v} \rangle = \sum_{w \neq v} 
\left( m_{vw} x_{w} - m_{wv} x_{v} \right) (\delta t) + \ldots\,,
\label{first_jm_1}
\end{equation}
where the dots indicate higher orders in $\delta t$. Therefore, from
Eq.~(\ref{jump_mom_def1}), 
$\alpha_{v} (\vec{x}) = \sum_{w \neq v}( m_{vw} x_{w} - m_{wv} x_{v})$. To
find the second jump moment, we need to consider 
$\langle \delta x_{v} (t) \delta x_{w} (t) \rangle$, but from 
Eq.~(\ref{delta_x_1}) we see that this is already ${\cal O} (\lambda^{2})$,
that is, ${\cal O} (\delta t)$. Therefore any terms in the matrix $M$ which
vanish as $\delta t \to 0$ do not contribute at this order. Since all 
off-diagonal entries and diagonal entries apart from 1, are of this form, $M$ 
may be replaced by the unit matrix everywhere in this second order term, i.e., 
any bias can be neglected. Using Eq.~(\ref{delta_x_1}) and Eq.~(\ref{navg_1}) 
with $\vec{x}^{\,\prime}$ replaced by $\vec{x}$, we obtain
\begin{equation}
\langle \delta x_{v} \delta x_{w} \rangle = \frac{1}{T^2} (\delta t) 
\left( \avg{n_{v} n_{w}} - \avg{n_{v}} \avg{n_{w}} \right) + \ldots\,.
\label{second_jm_1}
\end{equation}
Now the variance of the multinomial distribution is given by
\begin{multline}
\label{nnavg_1}
\avg{n_{v} n_{w}} - \avg{n_{v}} \avg{n_{w}} =
\left\{ \begin{array}{ll}
T x_{v}^\prime(1-x_{v}^\prime) & v=w,\\[1ex]
-T x_{v}^\prime x_{w}^\prime & v \ne w,
\end{array} \right.
\end{multline}
and so once again replacing $\vec{x}^{\,\prime}$ by $\vec{x}$ and using the
definition of the jump moment (\ref{jump_mom_def2}), we obtain 
Eq.~(\ref{second_jump_mom_2}). All higher jump moments vanish, since from
Eq.~(\ref{delta_x_1}) we see that the third and higher moments of 
$\delta \vec{x}$ are at least ${\cal O} (\lambda)^{3}$, that is, at least
${\cal O} (\delta t)^{3/2}$. Therefore the Kramers-Moyal expansion is truncated
at second order and we obtain the Fokker-Planck equation
\begin{eqnarray}
\label{FPE_1}
\frac{\partial P(\vec{x}, t)}{\partial t} &=& 
- \sum_{v=1}^{V-1} \frac{\partial}{\partial x_{v}} \sum_{w \neq v} 
\left( m_{vw} x_{w} - m_{wv} x_{v} \right) P(\vec{x}, t) \nonumber \\
&+& \frac{1}{2T} \sum_{v,w} \frac{\partial^{2} }{\partial x_v \partial x_w}
(x_v \delta_{v,w} - x_v x_w) P(\vec{x}, t)\,. \nonumber \\
\end{eqnarray}
The derivation in the case of the full model with $N$ speakers follows similar 
lines. Here $X(t) = ( \vec{x}_1(t), \ldots, \vec{x}_N(t))$ is an $N(V-1)$ 
dimensional grammar variable whose components we have written as $x_{i v}$. 
It is sometimes convenient to replace the two labels $\{ i, v \}$ by the single
one $I$ with $I=1,\ldots,N(V-1)$. Then Eqs.~(\ref{KM_1})-(\ref{jump_mom_def2})
in the derivation of the one-speaker case can be taken over by replacing $v$ 
and $w$ by $I=\{ v,i \}$ and $J=\{ w,j \}$ respectively. In the full utterance 
selection model, there is randomness both in the choice of speakers that 
interact in the interval $\delta t$ following time $t$ and in the tokens they 
produce.  

The jump moments are derived from averages of products of the quantity 
$\delta x_{I} = x_{I}(t+\delta t) - x_{I} (t)$. From (\ref{update}) we find
the analog of the one-speaker result (\ref{delta_x_1}) to be
\begin{equation}
\label{deltax}
\delta x_{i v} = \frac{\lambda}{1+\lambda(1+H_{i j})} \left[
\frac{n_{i v}}{T} - x_{i v} + H_{i j} \left( \frac{n_{j v}}{T} -
x_{i v} \right) \right]
\end{equation}
for a speaker $i$ given that speakers $i$ and $j$ have been already be
chosen as the interacting pair in the time-step at $t$.

The mean change in the grammar variable $\avg{\delta x_{iv}}$ can then
be determined by knowing that the mean of the multinomial distribution
(\ref{multinomial}) is simply
\begin{equation}
\label{navg}
\avg{n_{i v}} = T x_{i v}^\prime \;.
\end{equation}
Then
\begin{eqnarray}
\avg{\delta x_{i v}} &=& \frac{\lambda}{1+\lambda(1+H_{i j})} \left[
x_{i v}^\prime - x_{i v} + H_{i j} \left( x_{j v}^\prime - x_{i v}
\right) \right] \nonumber \\
\label{xmean}
&=& \lambda \left[ \sum_{w \ne v} \left(M_{vw} x_{iw} - M_{wv}
x_{iv}\right) + H_{ij} \left( x_{jv} - x_{iv}\right) \right] \nonumber\\
&& {} +  \Order(\lambda H M, \lambda^2 H, \lambda^2 M)
\end{eqnarray}
in which the second line was arrived from the first by using
(\ref{lintrans}).  Similarly, from the variance of the multinomial
distribution
\begin{multline}
\label{nnavg}
\avg{n_{i v} n_{j w}} - \avg{n_{i v}} \avg{n_{j w}} = \\
\left\{ \begin{array}{ll}
T x_{i v}^\prime(1-x_{i v}^\prime) & v=w, i=j\\[1ex]
-T x_{i v}^\prime x_{i w}^\prime & v \ne w, i=j\\[1ex]
0 & i \ne j
\end{array} \right.
\end{multline}
one finds
\begin{equation}
\label{xvar}
\avg{\delta x_{iv} \delta x_{jw}} = \frac{\lambda^2}{T} \left( x_{iv}
\delta_{v,w} - x_{iv} x_{iw} \right) + \Order(\lambda^2 H, \lambda^2
M, \lambda^3)
\end{equation}
if $i=j$ and $\avg{\delta x_{iv} \delta x_{jw}} = 0$ otherwise.

In order to have both a deterministic and stochastic part to the
Fokker-Planck equation, we need both $\avg{\delta x_{iv}}$ and $\avg{
\delta x_{iv} \delta x_{iw} }$ to be proportional to $\delta t$ in the 
limit $\delta t \to 0$. One can verify that the only way this can be 
arranged is if one rescales the variables as in 
Eqs.~(\ref{rescale1})--(\ref{rescale3}), a choice which was motivated in more 
detail in Section \ref{continuous}.  Then, only the leading terms in
Eqs.~(\ref{xmean}) and (\ref{xvar}) remain in when one takes the limit
$\delta t \to 0$ in Eqs.~(\ref{jump_mom_def1}) and (\ref{jump_mom_def2}).
Furthermore, all higher jump moments vanish, as also discussed in Section
\ref{continuous}, and the sum in Eq.~(\ref{KM_1}) terminates at the second 
moment. After substituting the jump moments into (\ref{KM_1}) and
averaging over all possible pairs of speakers, weighted by the
interaction probabilities $G_{ij}$, one finally arrives at the
Fokker-Planck equation given in the main text, Eq.~(\ref{driftfpe}).

\end{document}